# "Predicting Pathogenicity Of nsSNPs Associated With Rb1- An In Silico Approach"

**Anum Munir[1]**

[1]Department of Computer Science/Information Technology/Software Engineering NUML University of Modern Languages Islamabad Multan Campus 60000, Pakistan

Corresponding Author: Anum Munir. Email: anum.munir@numl.edu.pk

**Abstract:** Single nucleotide polymorphisms (SNPs) are variations at specific locations in DNA. Sequence responsible for marking genes associated with diseases or tracking inherited diseases within The family. These variations in the **Rb1** gene can cause Retinoblastoma and cancer in the retina Of one eye or both, Osteosarcoma, Melanoma, Leukemias, Lungs, and Breast cancer. First of all, The SNP database hosted by NCBI was used to extract some principal data. The association of Rb1 to Other genes were analyzed by GeneMANIA. Ten different computational tools, i.e., SIFT,Polyphen-2, I-Mutant 3.0, PROVEAN, SNAP2, PHD-SNP, PMut, SNPs&GO were used for the screening of damaging SNP for the estimation of conserved regions of amino acids Consurf Server was used for the evaluation of the structural stability of both native and mutant proteins, Project Hope was used to examine the structural effects of mutant protein . GeneMANIA predicted that RB1 Gene was expected to have a strong association with 20 other genes i.e. CCND1 and RBP2 etc. As per data retrieved from dbSNP hosted by NCBI,the Rb1 gene probed in this study carried a total of 36,358 SNPs. 345 were found in 3'UTR, 65 in 5'UTR, and 34,543 were found in the intron region. 844 were coding SNPs, and out of 844, 199 were synonymous And 450 were non-synonymous, including 425 missense, five nonsense, and 20 frameshift mutations. And remaining all are other types of SNPs. We took 425 missense SNPs for our investigation. A total of 17 mutations i.e. D332G, R445Q, E492V, P515T, W516G, V531G, E533K, E539K, M558R,W563G, L657Q, A658T, R661Q, D697H, D697E, P796L and R798W were predicted to have Damaging effects on structure and function of Rb1 protein..

## 1 Introduction

The work of this research has been presented in two parts as this is based on bioinformatics.The Foremost goal of the first part is to fetch data from the database and the purpose of the second part is to Extract accurate results using different bioinformatics tools by utilizing the same data taken from the database**. (first part of this study**) Rb1 stands for Retinoblastoma, which is cancer in the eye's retina, and it often develops in early Childhood. It may produce sporadically or genetically but is much more hostile if untreated. Here we Discuss the in-a-silico way out of treating this genetic disorder through a number of Computational tools. RB1 is a tumor suppressor gene found on chromosome no 13 that is responsible for the regulation of cell growth and prevents the cell from excessive growth in an uncontrolled way by Hindering cell cycle progression until cell division and when the cell is about to divide, Rb1 is phosphorylated to pRb, which leads to the inactivation of the Retinoblastoma protein activity (1). This phase permits cells to enter into the cell cycle state, which may lead to the Mutation of this gene (2)**.** However, if Rb1 is activated chronically, then it redirects to the decline of required DNA replication factors, and after 72-96 hours of this chronicle activation then, all targeted Proteins portray declined factors of DNA replication, and sometimes, it may lead to a grave condition in which hindrance of DNA replication in cells takes place (3). Rb1 is a component of "pocket protein family". Retinoblastoma protein **(RB)**, Retinoblastoma-like protein 1**(p107),** and Retinoblastoma-like protein 2**(p130)** is included in the pocket protein family**.** All three members Of this family are able the binding its functions to at least 100 further proteins, or shortly we can say that Rb1 is a multitasking protein with many Phosphorylation and binding sites, especially with E2F family. (4)

As this is a matter of fact that the total number of nucleotides in the Human Genome is 32,00,000,000, and they are split up into 24 linear molecules. As our study is on the RB1 gene Located on chromosome# 13, and the study reveals that chromosome # 13 has almost 114 Million base pairs comprise 3.5%-4.0% of the total genome in cells**.** Besides it, there are also,some variations in these nucleotides are called **"**single nucleotide polymorphisms (SNPs)" which Come about at the frequency of 1% of the total nucleotides, which may differ. For different populations or shortly we can say that SNPs occur almost once in every 1000 Nucleotides. This way, 4-5 million SNPs exist in the human genome. According to one study, scientists found that almost there are 100 million SNPs are there in the World. These variations are found between genes in DNA (5). SNPs can function independently but In some cases, SNPs work in groups with other SNPs and become the source of major.The disease can be observed in osteoporosis (6). According to one study, nsSNPs are the root Cause because more than 50% of mutations are connected with genetic diseases (7). However, there are Different types of mutations. Some mutations are not so severe. They result in mild phenotype(8).Several hundred thousand mutations have been revealed that affect the proper functionality of the Rb1 gene (9). Recent studies have revealed that the victim of Retinoblastoma is One out of 16,000-18,000 live births that leads to 9,000 new emerging cases annually worldwide(10). The common symptoms are the abnormal look of the pupil and leukocoria, which is also known as "amaurotic cat's eye reflex" or "leukocoria". Leuko means "white" and coria means "pupil". This Abnormal whiteness can be noticed in photographs captured with a flash and dim light. Other symptoms can be crossed eyes or that condition of visions in which eyes can't see or point in the same direction. This condition is called Strabismus which causes some variations in the color of the iris, soreness, redness, swollen eyelids, and poor vision (11). Retinoblastoma is of two types which are unilateral Retinoblastoma and bilateral Retinoblastoma.(12). Former infects only one eye in the majority and its diagnosis age is 24 months but later Infects both eyes in one out of three children whose diagnosis age is 12 months. Children who are having any type of Retinoblastoma, they may lose their vision, and sometimes they need Removal of eye or eyes (13). According to researchers, there are two forms of Retinoblastoma. One is hereditary Retinoblastoma. And the other is non-hereditary Retinoblastoma. People are having this form of Retinoblastoma pass these variations to their next generation, most probably as hereditary Retinoblastoma autosomal Dominant, which means that if one copy of the mutated gene is present in one of the parents, then here is a 50% chance of inheriting that gene from each child in next-generation and when this mutated gene will pass to the next generation, and then victims will have numerous tumors in both eyes. On the other Hand, the other form of cancer is non-hereditary, which means that Rb1 cells can't be transferred to the next generation. In this form of Retinoblastoma, affected individuals are reproduced with two regular copies of the Rb1 gene but with time in early childhood, these two copies of the Rb1 gene exhibit some mutations but these mutations can't beTransferred to the next generation (14). Generally, in this form, children have one tumor in one eye.It is impossible to know whether the person having Retinoblastoma in one eye is hereditary or Non-hereditary in nature, but with proper "**genetic testing**", It is easy to know the form of Retinoblastoma (15).In addition to this, when Rb1 is imparted with genetic mutations, then it spreads in all of the body cells called "hereditary or germinal retinoblastoma ."This form of Rb1 is much dangerous as it causes cancers in both eyes or outside the eye (16).

Mutations in Rb1 are also the root cause of **"Pineoblastoma"** (cancer in the pineal gland of the brain) (17). The genetic mutation in Rb1 is also responsible for breast cancer, lung cancer, bone cancer called Osteosarcoma. Mutations are also responsible for cancer of soft tissues called. "sarcomas" typically identifies at approximately 10-20 years, mainly in Rb survivors(18). Mutation in Rb1 is also responsible for a hostile form of skin cancer called Melanoma which arises typically in the skin, but sometimes, in most rare cases, it also comes up in the mouth, intestines and eyes (19). As far as the development of melanomas is concerned, almost 25% of Melanomas originate from the simple mole (20). Somatic Rb1 mutations have been reported in cancers of blood-forming cells called leukemias (21). The scope of this study is to provide suitably Personalized medication to every individual after a detailed analysis of the Person's Genome. However, it is difficult or a hindrance to detect functional SNPs in any particular gene using Laboratory tools and techniques But due to the enhancement and advancement in in-silico,we can now do all such detections of functional SNPs without any extensive laboratory Work. The foremost goal of this study is to look into genetic variations of SNPs and their possible impact on the structure and function of the Rb1 gene with the help of different Computational tools and techniques.

## 2 Materials and Methods

### 2.1 Data Gathering:

All the required SNP particulars of the human Rb1 gene, i.e., "rsids, Protein accession Number=NP_000312.2, mRNA accession Number=NM_000498.3, and residue changes" have been extracted From National Center for Biotechnology Information (NCBI), which is SNP's Public database i.e.dbSNP **https://www.ncbi.nlm.nih.gov/**).

Out of 844 coding SNPs, 425 missense SNPs were taken Into account for further analysis using different bioinformatics tools. Although a division of Coding SNPs at different regions is shown in Figure 3-1

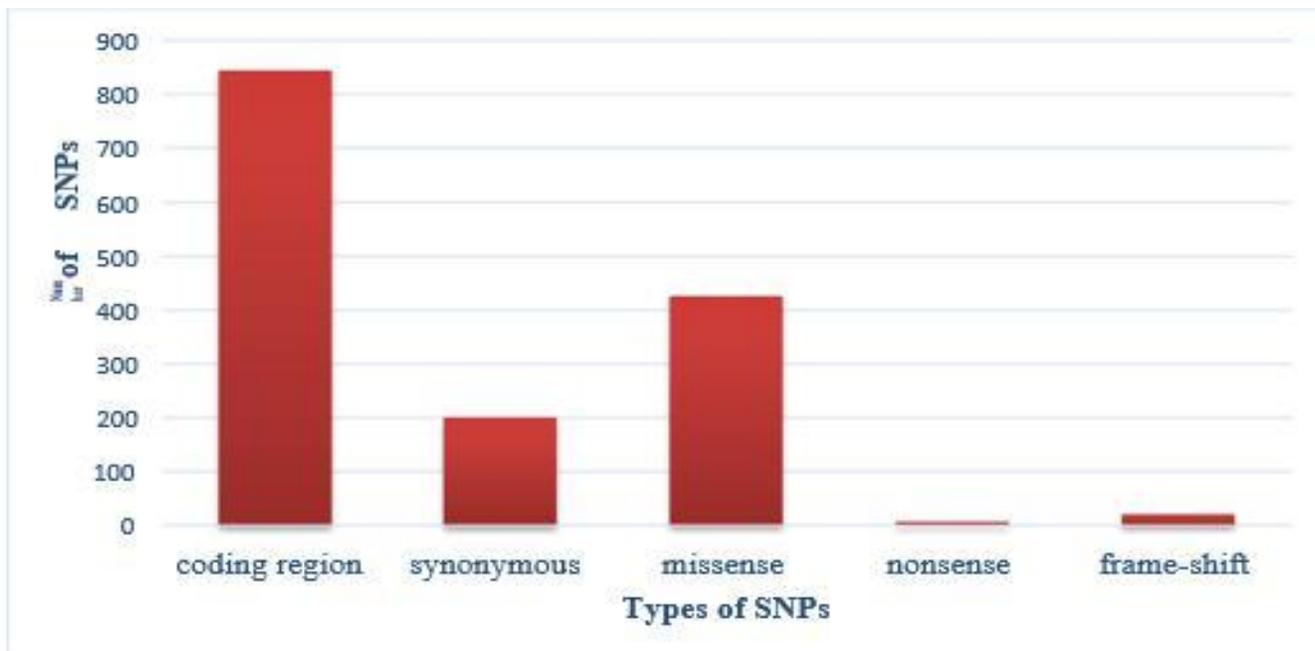

**Fig 3. Graphical Portray regarding the division of coding SNPs for Rb1 Gene(as per NCBI)**

## 2.2 Scanning the association of Rb1 with other Genes using GeneMANIA:

GeneMANIA is a software that contains 597,392,988 associations of 163,599 genes that tell the function, i.e., genetic interactions, physical interactions, pathways, co-localization, and protein domain similarity of any gene as well as its association with other genes (26).

GeneMANIA has an integrated plugin that uses an extensive database containing data from multiple.

Organisms. GeneMANIA is available at the following link; **https://genemania.org/.**

### 2.3 The anticipation of Damaging and Tolerated nsSNPs, as well as their functional effects by SIFT:

**"Sorting Intolerant from tolerant (SIFT)"** algorithm being utilized, which anticipates in by sequence homology that whether the Substitution of an amino acid affects protein Function or not as well as is also able to predict physical properties and Chemical properties of amino acids. All the above data was taken from NCBI (https://www.ncbi.nlm.nih.gov/). By SIFT, we anticipated the deleterious and tolerant impact of non-synonymous SNPs found in the **coding region** with the help of tolerance Index (TI) score. Based on this score, we can anticipate tolerated and deleterious amino acid substitutions. The threshold for tolerance ranges from 0.0 to 1.0 while deleterious is < 0.05 (27). SIFT analysis was able to find homologous sequences based on algorithms by using different databases, i.e., Swiss-Prot and TrEMBL, by choosing "Median Conservation Sequence Score 3.00. Sift can be found at the following link; "(https://sift.bii.a-star.edu.sg/)."

### 2.4 The anticipation of functional impacts of nsSNPs using PolyPhen-2:

PolyPhen-2 (Phenotyping Polymorphism) software automatically predicts the possible impact of amino acid substitution on the structure and function of human protein (28). Polyphen-2 detects "3D protein structures" and "contact information" of amino acids to compute position-specific independent counts *"**PSIC** "*scores for both variants of amino acids and then computes PSIC score difference. The score difference is directly proportional to the functional effect means the higher the score of "**PSIC**", the higher will be the functional effect for each.

Substitution of amino acid. The scores of the polyphen-2 range from **0.0 -1.0.**

Variants with scores from **0.0 to 0.15** are anticipated to be benign.

   Variants having scores ranging from **0.15-1.0** are anticipated to be possibly damaging

Variants with scores ranging from **0.85-1.0** are indeed anticipated to be damaging [52].

Polyphen-2 is available at the following link; **http://genetics.bwh.harvard.edu/pph2/.**

### 2.5 The anticipation of protein Stability using I-Mutant 3.0:

I-Mutant 3.0 anticipates variation in protein stability based on neural networks and Support Vector Machine **(SVM)**. It has also been analyzed that I-mutant can do better than other tools.

"Distance-Scaled, Finite Ideal Gas Reference (DFIRE)," "FoldX," and "Prediction of Protein Mutant Stability Changes (PoPMuSiC)*"* (29). In I-Mutant 3.0 wild type of free energy was subtracted from the mutant accessible point to get "free energy change (ΔΔG)." + ive sign of (ΔΔG) means the highest stability of protein. Conversely, the –ive sign of (ΔΔG) indicates decreased protein stability. Shortly, We can say that if (ΔΔG) > 0, this means increased protein stability, and if (ΔΔG) < 0, it means decreased protein Stability [54].

I-Mutant 3.0 is found at the following link;

**"http://gpcr2.biocomp.unibo.it/cgi/predictors/I-Mutant3.0/I-Mutant3.0.cgi"**

## 2.6 Identification of functional nsSNPs using PROVEAN:

**PROVEAN ("Protein Variation Effect Analyzer")** is a fast computational approach that identifies that nsSNPs which are functionally important, means it anticipates whether substitution of amino acid affects the **"protein's biological function"** or not. Scores less than -2.5 means deleterious SNPs, and greater than -2.5 are known as neutral SNPs (30)**.**

PROV EAN is found at the following link;

**("http://PROVEAN.jcvi.org/seq_submit.php")**

## 2.7 Validation to anticipate the functional impacts of sequence variants using SNAP2:

**SNAP2** is the bioinformatics tool used to anticipate the functional impacts of sequence variants or mutations of the amino acid (31). It is a sort of trained classifier based on the concept of machine learning devices are called neural networks. It plays a vital role in differentiating between effected and Neural nsSNPs by just taking simple data into account.

Its **Visual representation** is somehow technical, as shown in **Fig 4.**

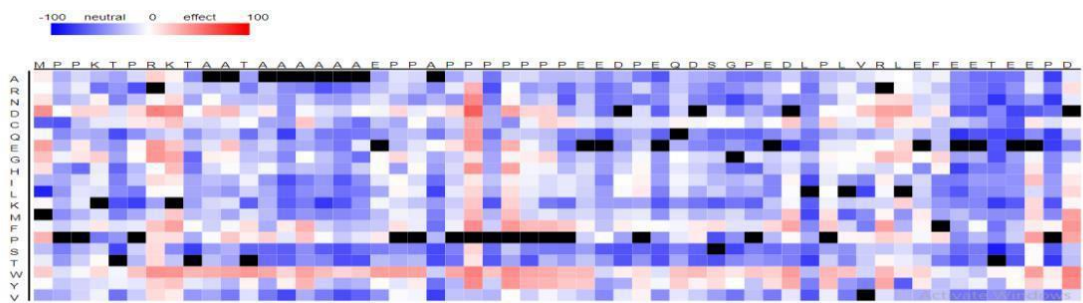

**Fig 4. Heatmap Visual Representation**

This visual representation has the following criteria;

**Dark red** tiny blocks indicate score > 50 means a Strong Signal yet affected. **White** tiny blocks indicate -50<score<50 means weak signal.

**Green** tiny blocks mean score <-50 means strong signal yet neutral.

SNAP2 is available at the following link;

**https://www.rostlab.org/services/snap/**

### 2.8 Predicting diseased mutations Using PHD-SNP:

PhD-SNP(*"Predictor of human deleterious single nucleotide polymorphism"*) is a bioinformatics software that is based on the **Support vector machine (SVM)**. This software is being utilized to sort out a diseased or neutral substitution of amino acid out of those deleterious nsSNPs which are being declared deleterious by all the above tools, i.e., SIFT, Polyphen-2, I-Mutant 3.0, PROVEAN, and SNAP2. The primary rationale behind using this tool is basically the verification to get accurate results (32),

*"PhD-SNP"* is found at the following link;

"**http://snps.biofold.org/phd-snp/phd-snp.html**"

### 2.9 Predicting diseased mutations Using PMut:

But is a bioinformatics web server used to find out disease-causing mutations. It isn't very objective.on Neural Networks for the processing of information (33) but provides simple Predictions in the form of neutral or disease mutations as well as prediction Scores.

PMut is available on the following link**;**

**http://mmb.irbbarcelona.org/PMut/**

### 2.10 Predicting diseased mutations Using SNPs&GO:

It is a bioinformatics tool based on a simple vector machine (SVM) to find mutations Causing disease. After giving input, the tool will process the information and provide the output in the form of Neural or disease mutations, as well as give the Reliability index RI. If the RI value is shown by the tool is greater than 5, indicating that this is a disease-related protein mutation (34)**.**

SNPs&GO is available on the following link;

**http://snps.biofold.org/snps-and-go/snps-and-go.html**

**2.11 Evolutionary conservation analysis of nsSNPs:**

The Consurf Server is another bioinformatics tool used to estimate the evolutionary conservation score of amino acid's position to maintain the function and structure of the Rb1 gene (35). Evolutionary relationships estimate consurf Scores. The conservation score Scale is mentioned in the results in a diagram and table (obtained from UniRef-90).The accuracy of the conservation score has significantly improved due to the empirical Bayesian method but provided that no sequences for calculation should be less (36).

Consurf Server is available on the following link;

**https://consurf.tau.ac.il/**

**2.12 Predicting the impact of nsSNPs on 3D Protein Structure:**

Another bioinformatics tool is Project Hope which is a user-friendly web service tool used to Collect 3D protein structures after doing calculations on 3D protein coordinates, taking sequences of annotations on the database of **"UniProt"** and getting estimations by **"DAS services**." (37)

*"Project hope"* is found at the following link;

**(http://www.cmbi.ru.nl/hope/)**

Below is the graphical representation of the overall methodology being utilized in this study

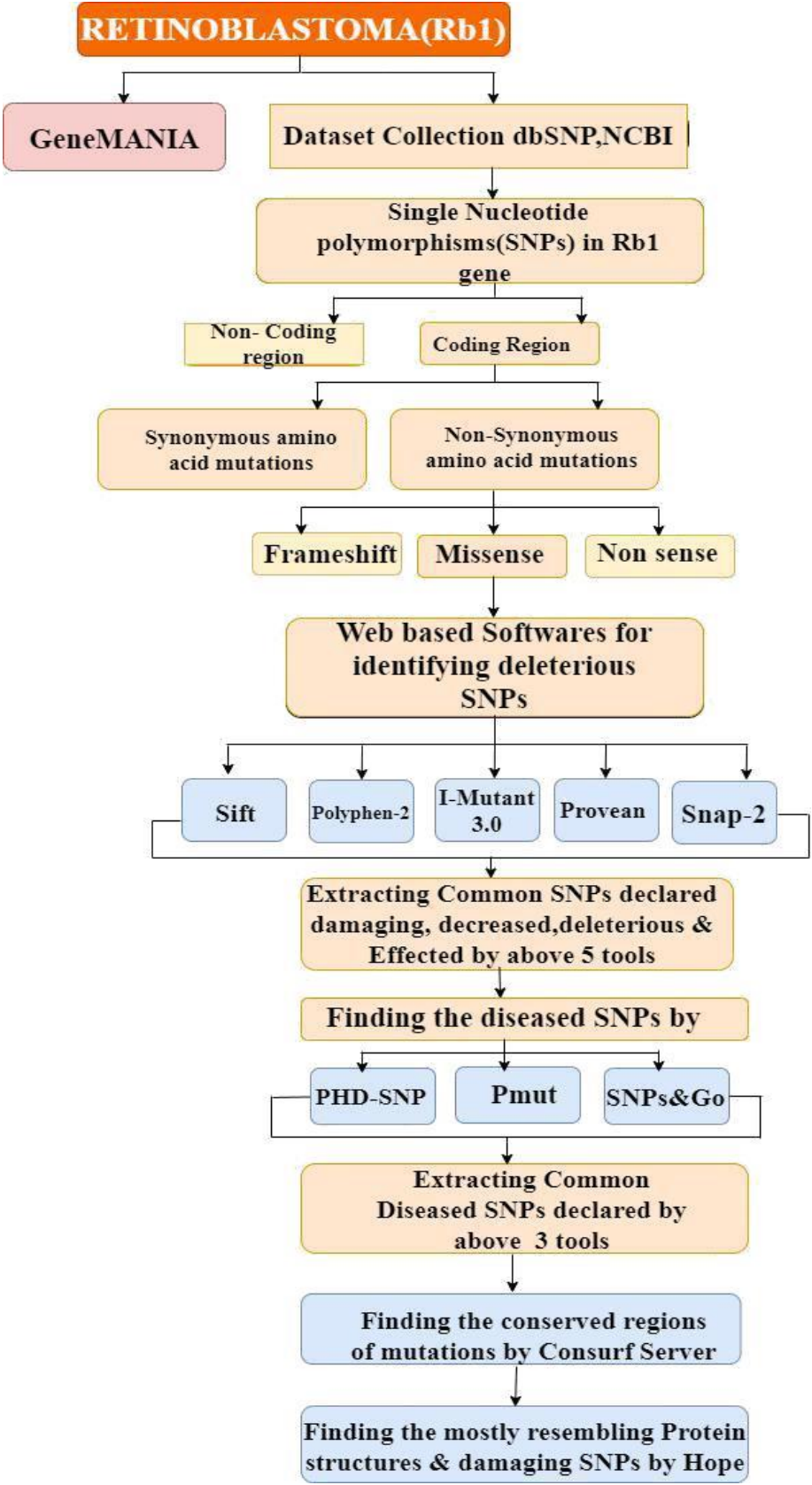

**Fig 5. Workflow of in silico methods used in this Study**

# 3. Results

### 3.1 Result of GeneMANIA:

According to the GeneMANIA software, Rb1 is associated with 20 other genes. Among these 20, CCND1 is strongly associated with Rb1, the sub-component of "holoenzyme." Different parameters i.e., physical interactions, genetic interactions, and co-expression of the Rb1 gene with

Other genes are shown in **Table 1** and **Fig 1.**

### Table 1. Gene Description Rank Using GeneMANIA

| Sr # | Genes | Description | Rank |
|---|---|---|---|
| 1. | RB1 | RB transcriptional corepressor 1 | N/A |
| 2. | CCND1 | cyclin D1 | 1 |
| 3. | RBP2 | retinol-binding protein 2 | 2 |
| 4. | ATF7 | activating transcription factor 7 | 3 |
| 5. | CBX4 | chromo box 4 | 4 |
| 6. | TAF1 | TATA-box binding protein associated factor 1 | 5 |
| 7. | GTF3C2 | general transcription factor IIIC subunit 2 | 6 |
| 8. | ID2 | inhibitor of DNA binding 2, HLH protein | 7 |
| 9. | ELF1 | E74, like ETS transcription factor 1 | 8 |
| 10. | CDK4 | cyclin-dependent kinase 4 | 9 |
| 11. | PAX3 | paired box 3 | 10 |
| 12. | BRF1 | BRF1, RNA polymerase III transcription initiation factor 90 kDa subunit | 11 |
| 13. | RBL1 | RB transcriptional corepressor like 1 | 12 |
| 14. | RBL2 | RB transcriptional corepressor like 2 | 13 |
| 15. | PURA | purine-rich element binding protein A | 14 |

| 16. | PPP2R3B | protein phosphatase 2 regulatory subunit B"beta [Source:HGNC Symbol; Acc: HGNC:13417] | 15 |
| 17. | SUV39H1 | suppressor of variegation 3-9 homolog 1 | 16 |
| 18. | IRF3 | interferon regulatory factor 3 | 17 |
| 19. | PAX5 | paired box 5 | 18 |
| 20. | DNMT1 | DNA (cytosine-5-)-methyltransferase 1 | 19 |
| 21. | GTF2H1 | general transcription factor IIH subunit 1 | 20 |

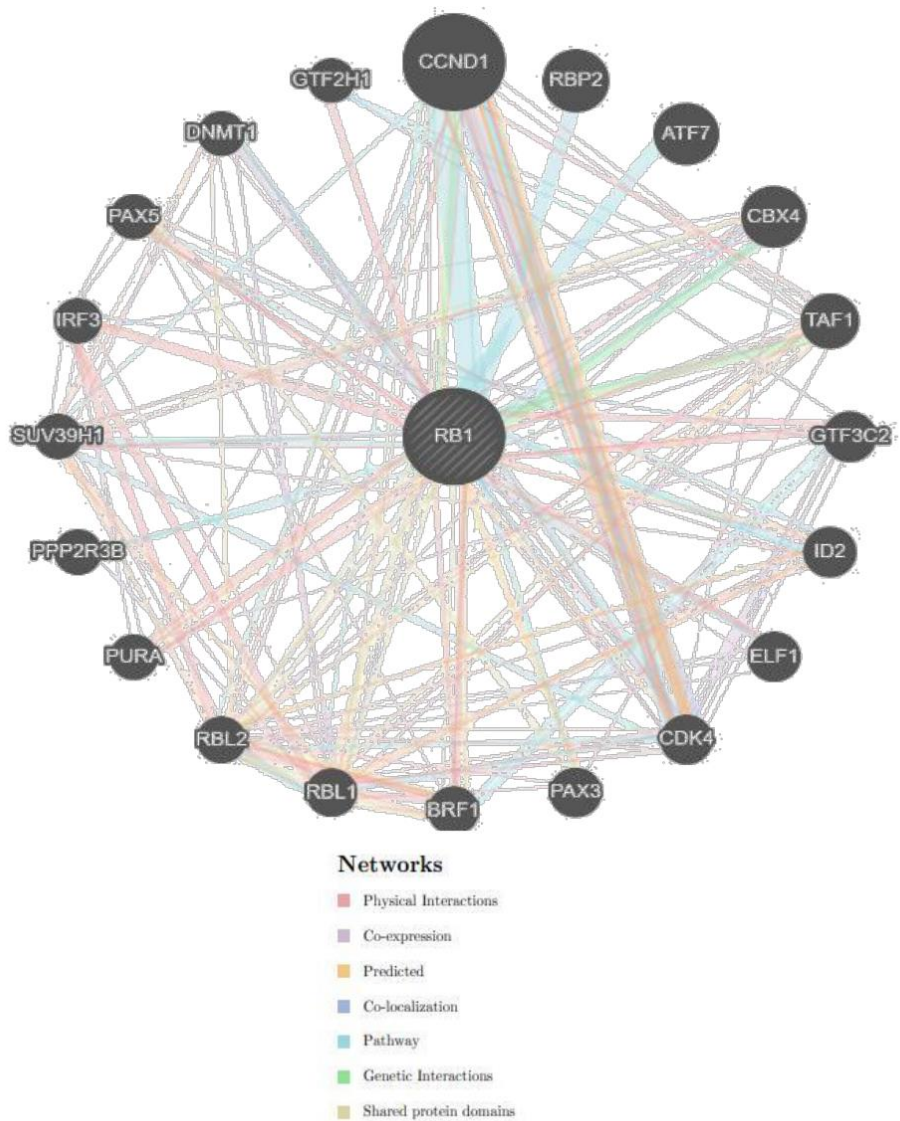

**Fig 1. Results of GeneMANIA for Rb1 Gene**

A brief Summary of all the tools used in this study to detect the pathogenicity of nsSNPs is given in **Table 2**

**Table 2. Summary**

| Sr # | Tools | Results |
|---|---|---|
| **1.** | Sift | Total SNPs=**425**<br>Damaging=**139**<br>Tolerated=**276**<br>N/A=**7**<br>Not secured=**3** |
| **2.** | Polyphen-2 | Total SNPs=**425**<br>Possibly damaging=**93**<br>Probably Damaging=**156**<br>Benign=**176** |
| **3.** | I-Mutant3.0 | Total SNPs=**425**<br>Decreased =**325**<br>Increased=**100** |
| **4.** | PROVEN | Total SNPs=**425**<br>Deleterious=**89**<br>Neutral=**336** |
| **5.** | SNAP2 | Total SNPs=**425**<br>Effect=**200**<br>Normal=**225** |
| **6.** | PHD-SNP | Total Common damaging SNPs=**40**<br>Diseased=**26**<br>Neutral=**14** |
| **7.** | Pmut | Total Common damaging SNPs=**40**<br>Diseased=**24**<br>Neutral=**16** |
| **8.** | SNP& GO | Total Common damaging SNPs=**40** |
|     |                | Diseased=**40**                                                                   |
|     |                | Neutral=**0**                                                                     |
| 9.  | Consurf Server | Total Common damaging SNPs=**21**<br>Conserved=**19**<br>Non-Conserved=**2**      |
| 10. | HOPE           | Total Common damaging SNPs=**21**<br>Damaging=**17**<br>Non –Damaging=**04**      |

### 3.2 SIFT Results:

Out of all coding SNPs, only missense SNPs were taken into account, as these nsSNPs directly affect the structure and function of the protein. First, to find out whether the mutations affect the part of the protein, a total of 425 missenses were submitted into SIFT. Out of 425, 139 were declared as damaging nsSNPs.

### 3.3 Polyphen-2 Results:

Polyphen-2 is used to analyze the possible effect of amino acid substitution on a protein's structure and function. When 425 nsSNPs were submitted into polyphen-2,93 were declared as Possibly damaging,156 as Probably Damaging, and 176 were Benign.

### 3.4 Mutant 3.0 Results:

I-Mutant 3.0 is used to automatically analyze changes in Protein stability upon mutations on specific points. I-Mutant predicted 325nsSNPs as decreased stability when all 425 nsSNPs were submitted into I-Mutant 3.0.

### 3.5 PROVEAN Results:

**PROVEAN Server** to further forecast whether amino acid substitution affects the biological function of the protein or not. PROVEAN declared 89 nsSNPs as deleterious when all 425 nsSNPs were submitted to it.

### 3.6 SNAP2 Results:

After this, **SNAP2** was used to determine the impact of mutations on the protein's function. SNAP2 declared 200 nsSNPs as affected out of 425.

### 3.7 Results of Comparison Method:

To go for the best, the **comparison method** is used to compare the results of all the above tools. **Table 3** shows the **40** common **deleterious** nsSNPs declared by **SIFT Polyphen-2, I-Mutant 3.0, PROVEAN, and SNAP2.**

# Table 3 : 40 COMMON nsSNPs DECLARED DAMAGED BY SIFT, Polyphen-2, I-Mutant3.0, PROVEAN & SNAP2

| Sr# | Rsid's | AA Change | SiftPrediction | Score | Polyphen-2 Prediction | Score | I-Mutant 3.0 Prediction | DDG value | PROVEAN Prediction | Score | SNAP2 Prediction | Score |
|---|---|---|---|---|---|---|---|---|---|---|---|---|
| 1. | rs983885759 | L120Q | Damaging | 0 | Probably Damaging | 1 | Decrease | -1.44 | Deleterious | -3.968 | Effect | 69 |
| 2. | rs749495284 | F131L | Damaging | 0 | Probably Damaging | 0.999 | Decrease | -1.09 | Deleterious | -3.786 | Effect | 67 |
| 3. | rs766849295 | Y155C | Damaging | 0 | Probably Damaging | 1 | Decrease | -0.72 | Deleterious | -5.529 | Effect | 53 |
| 4. | rs1034616967 | D156Y | Damaging | 0.03 | Probably Damaging | 0.93 | Decrease | -0.56 | Deleterious | -2.689 | Effect | 40 |
| 5. | rs141366046 | V190G | Damaging | 0.01 | Probably Damaging | 0.695 | Decrease | -2.74 | Deleterious | -2.575 | Effect | 69 |
| 6. | rs1267000381 | V213M | Damaging | 0 | Probably Damaging | 1 | Decrease | -0.28 | Deleterious | -2.6 | Effect | 79 |
| 7. | rs1273219762 | C283Y | Damaging | 0.01 | Probably Damaging | 0.949 | Decrease | -1.79 | Deleterious | -4.455 | Effect | 61 |
| 8. | rs765678030 | E323G | Damaging | 0.01 | Probably Damaging | 0.969 | Decrease | -1.63 | Deleterious | -3.883 | Effect | 50 |
| 9. | rs868847993 | Y325H | Damaging | 0 | Probably Damaging | 0.605 | Decrease | -0.82 | Deleterious | -3.158 | Effect | 74 |
| 10. | rs763377384 | D330N | Damaging | 0.03 | Probably Damaging | 0.605 | Decrease | -2.08 | Deleterious | -3.695 | Effect | 67 |
| 11. | rs763184576 | D332G | Damaging | 0.03 | Probably Damaging | 0.999 | Decrease | -1.05 | Deleterious | -5.239 | Effect | 73 |
| 12. | rs748635133 | G423E | Damaging | 0.05 | Probably Damaging | 0.998 | Decrease | -1 | Deleterious | -4.381 | Effect | 69 |
| 13. | rs747509282 | R445Q | Damaging | 0 | Probably Damaging | 1 | Decrease | -0.4 | Deleterious | -2.985 | Effect | 69 |
| 14. | rs759079385 | R451C | Damaging | 0.01 | Probably Damaging | 1 | Decrease | -1.02 | Deleterious | -3.618 | Effect | 24 |
| 15. | rs771480219 | E492V | Damaging | 0 | Probably Damaging | 1 | Decrease | -1.62 | Deleterious | -6.529 | Effect | 79 |
| 16. | rs1158433317 | L512S | Damaging | 0.05 | Probably Damaging | 0.977 | Decrease | -0.79 | Deleterious | -2.526 | Effect | 67 |
| 17. | rs866664638 | P515T | Damaging | 0 | Probably Damaging | 1 | Decrease | -0.46 | Deleterious | -7.467 | Effect | 66 |
| 18. | rs138201027 | W516G | Damaging | 0 | Probably Damaging | 1 | Decrease | -1.62 | Deleterious | -10 | Effect | 81 |
| 19. | rs1331702695 | F526V | Damaging | 0.01 | Probably Damaging | 0.999 | Decrease | -0.62 | Deleterious | -5.333 | Effect | 67 |
| 20. | rs143324585 | V531G | Damaging | 0 | Probably Damaging | 1 | Decrease | -1.93 | Deleterious | -6.533 | Effect | 76 |
| 21. | rs1237070816 | E533K | Damaging | 0 | Probably Damaging | 1 | Decrease | -2.21 | Deleterious | -3.733 | Effect | 88 |

| #   | rsID          | Mutation | SIFT     | Score | PolyPhen-2        | Score | I-Mutant | DDG   | PROVEAN    | Score   | SNAP2  | Score |
|-----|---------------|----------|----------|-------|-------------------|-------|----------|-------|------------|---------|--------|-------|
| 22. | rs148379933   | E539K    | Damaging | 0.01  | Probably Damaging | 1     | Decrease | -2.35 | Deleterious | -3.3    | Effect | 79    |
| 23. | rs139494954   | M558R    | Damaging | 0     | Probably Damaging | 0.935 | Decrease | -1.61 | Deleterious | -4.5    | Effect | 46    |
| 24. | rs143400770   | L561P    | Damaging | 0.01  | Probably Damaging | 1     | Decrease | -0.02 | Deleterious | -5.332  | Effect | 75    |
| 25. | rs139500527   | W563G    | Damaging | 0     | Probably Damaging | 1     | Decrease | -2.32 | Deleterious | -12.133 | Effect | 88    |
| 26. | rs772068738   | P595L    | Damaging | 0.02  | Probably Damaging | 0.813 | Decrease | -0.49 | Deleterious | -3.915  | Effect | 35    |
| 27. | rs775051210   | S608C    | Damaging | 0.01  | Probably Damaging | 1     | Decrease | -2.04 | Deleterious | -3.279  | Effect | 26    |
| 28. | rs562956970   | L657Q    | Damaging | 0     | Probably Damaging | 0.939 | Decrease | -1.05 | Deleterious | -5.1    | Effect | 69    |
| 29. | rs202119986   | A658T    | Damaging | 0     | Probably Damaging | 1     | Decrease | -2    | Deleterious | -3.389  | Effect | 50    |
| 30. | rs750578651   | R661Q    | Damaging | 0     | Probably Damaging | 1     | Decrease | -0.93 | Deleterious | -3.581  | Effect | 84    |
| 31. | rs1172128543  | H686R    | Damaging | 0.01  | Probably Damaging | 1     | Decrease | -1.32 | Deleterious | -4.85   | Effect | 72    |
| 32. | rs775195256   | E693G    | Damaging | 0.01  | Probably Damaging | 1     | Decrease | -0.73 | Deleterious | -3.942  | Effect | 67    |
| 33. | rs1358369644  | D697H    | Damaging | 0.02  | Probably Damaging | 0.998 | Decrease | -1.11 | Deleterious | -4.5    | Effect | 72    |
| 34. | rs3092903     | D697E    | Damaging | 0.04  | Probably Damaging | 1     | Decrease | -0.72 | Deleterious | -2.796  | Effect | 16    |
| 35. | rs1363146373  | Y771C    | Damaging | 0.01  | Probably Damaging | 0.950 | Decrease | -1.81 | Deleterious | -4.943  | Effect | 59    |
| 36. | rs754507551   | P786S    | Damaging | 0.03  | Probably Damaging | 0.995 | Decrease | -0.75 | Deleterious | -2.688  | Effect | 14    |
| 37. | rs1467492987  | P789H    | Damaging | 0     | Probably Damaging | 0.997 | Decrease | -0.66 | Deleterious | -3.802  | Effect | 39    |
| 38. | rs1158706854  | P796L    | Damaging | 0     | Probably Damaging | 0.485 | Decrease | 0     | Deleterious | -5.221  | Effect | 19    |
| 39. | rs187912365   | R798W    | Damaging | 0     | Probably Damaging | 1     | Decrease | -1.21 | Deleterious | -3.986  | Effect | 75    |
| 40. | rs1394610552  | G836D    | Damaging | 0.05  | Probably Damaging | 1     | Decrease | -1.36 | Deleterious | -3.289  | Effect | 46    |

### 3.8 Results of PhD-SNP, PMut, and SNPs&GO:

To analyze whether these above 40 Common damaging SNPs are disease-causing or neutral, we will check them utilizing several tools, i.e., **PhD-SNP, PMut**, and **SNPs&GO.**

**"S2 Appendix"** shows the results of **"PHD-SNP, PMut, and SNPs&GO"** when these 40 joined deleterious nsSNPs (declared by Sift, Polyphen-2, I-Mutant 3.0, PROVEAN and SNAP2) were submitted to them.

**Table 4** shows that All three tools, i.e., PhD-SNP, PMut, and SNPs&GO display the result in the form of neutral and disease-causing mutations.

- PHD-SNP predicts 26 out of 40 mutations as Diseased and 14 as neutral mutations.

- PMut predicts 24 out of 40 as diseased and 16 as neutral mutations.

- SNPs&GO predicts 40 out of 40 mutations have disease effects.

That **21** nsSNPs which are bold, are the common diseased mutations declared by all the above three tools.

**Table 4 :LIST OF nsSNPs predicted as Diseased by PHD-SNP , PMut and SNPs&GO**

| Sr # | RSID's | AA Change | PHD-SNP Prediction | RI | PMut Prediction | Score | SNPs&GO Prediction | RI |
|---|---|---|---|---|---|---|---|---|
| 1. | rs983885759 | L120Q | **Disease** | **4** | **Disease** | **0.52** | **Disease** | **8** |
| 2. | rs749495284 | F131L | **Disease** | **4** | **Disease** | **0.79** | **Disease** | **9** |
| 3. | rs766849295 | Y155C | Neutral | 1 | Neutral | 0.49 | Disease | 9 |
| 4. | rs1034616967 | D156Y | Neutral | 7 | Neutral | 0.16 | Disease | 7 |
| 5. | rs141366046 | V190G | Neutral | 0 | Disease | 0.62 | Disease | 9 |
| 6. | rs1267000381 | V213M | Disease | 3 | Neutral | 0.43 | Disease | 9 |
| 7. | rs1273219762 | C283Y | **Disease** | **5** | **Disease** | **0.63** | **Disease** | **9** |
| 8. | rs765678030 | E323G | Neutral | 2 | Neutral | 0.48 | Disease | 7 |
| 9. | rs868847993 | Y325H | Neutral | 5 | Disease | 0.62 | Disease | 8 |
| 10. | rs763377384 | D330N | Neutral | 5 | Neutral | 0.14 | Disease | 7 |
| 11. | rs763184576 | D332G | **Disease** | **7** | **Disease** | **0.82** | **Disease** | **9** |
| 12. | rs748635133 | G423E | Disease | 3 | Neutral | 0.21 | Disease | 8 |
| 13. | rs747509282 | R445Q | **Disease** | **7** | **Disease** | **0.62** | **Disease** | **9** |
| 14. | rs759079385 | R451C | Disease | 4 | Neutral | 0.24 | Disease | 9 |
| 15. | rs771480219 | E492V | **Disease** | **7** | **Disease** | **0.82** | **Disease** | **10** |
| 16. | rs1158433317 | L512S | Neutral | 2 | Neutral | 0.34 | Disease | 9 |
| 17. | rs866664638 | P515T | **Disease** | **6** | **Disease** | **0.81** | **Disease** | **9** |
| 18. | rs138201027 | W516G | **Disease** | **9** | **Disease** | **0.7** | **Disease** | **10** |
| 19. | rs1331702695 | F526V | Disease | 9 | Neutral | 0.48 | Disease | 9 |
| 20. | rs143324585 | V531G | **Disease** | **8** | **Disease** | **0.77** | **Disease** | **10** |
| 21. | rs1237070816 | E533K | **Disease** | 8 | **Disease** | 0.82 | **Disease** | 10 |
| 22. | rs148379933 | E539K | **Disease** | 5 | **Disease** | 0.66 | **Disease** | 10 |
| 23. | rs139494954 | M558R | **Disease** | 7 | **Disease** | 0.62 | **Disease** | 6 |
| 24. | rs143400770 | L561P | **Disease** | 8 | **Disease** | 0.7 | **Disease** | 10 |
| 25. | rs139500527 | W563G | **Disease** | 9 | **Disease** | 0.82 | **Disease** | 9 |
| 26. | rs772068738 | P595L | Neutral | 4 | Neutral | 0.43 | Disease | 7 |
| 27. | rs775051210 | S608C | Neutral | 3 | Neutral | 0.47 | Disease | 8 |
| 28. | rs562956970 | L657Q | **Disease** | 9 | **Disease** | 0.73 | **Disease** | 10 |
| 29. | rs202119986 | A658T | **Disease** | 6 | **Disease** | 0.53 | **Disease** | 9 |
| 30. | rs750578651 | R661Q | **Disease** | 9 | **Disease** | 0.84 | **Disease** | 10 |
| 31. | rs1172128543 | H686R | Neutral | 1 | Neutral | 0.26 | Disease | 8 |
| 32. | rs775195256 | E693G | Disease | 3 | Neutral | 0.32 | Disease | 9 |
| 33. | rs1358369644 | D697H | **Disease** | 3 | **Disease** | 0.63 | **Disease** | 9 |
| 34. | rs3092903 | D697E | **Disease** | 1 | **Disease** | 0.56 | **Disease** | 8 |
| 35. | rs1363146373 | Y771C | Neutral | 0 | Disease | 0.6 | Disease | 9 |
| 36. | rs754507551 | P786S | Neutral | 3 | Neutral | 0.3 | Disease | 8 |
| 37. | rs1467492987 | P789H | Neutral | 0 | Neutral | 0.5 | Disease | 9 |
| 38. | rs1158706854 | P796L | **Disease** | 3 | **Disease** | 0.81 | **Disease** | 10 |
| 39. | rs187912365 | R798W | **Disease** | 3 | **Disease** | 0.67 | **Disease** | 9 |
| 40. | rs1394610552 | G836D | Neutral | 2 | Neutral | 0.5 | Disease | 9 |

### 3.9 Results of Consurf Server:

The role of evolutionary information is much more vital as it is used in finding out those mutations.

Which might give rise to detrimental effects on human health. For this, all that common diseased.

Mutations declared by PHD-SNP, PMut and SNPs&GO were submitted into the consurf web server in

order to estimate their evolutionary conservation scores because diseased nsSNPs located at

Conserved regions are incredibly high risk diseased to the protein's structure and function than nsSNPs located in non-conserved regions.

Evolutionary scores of above 21 diseased nsSNPs are being calculated from the consurf Server as Shown in **Fig**, and then their results were displayed in **"S3 Appendix."**

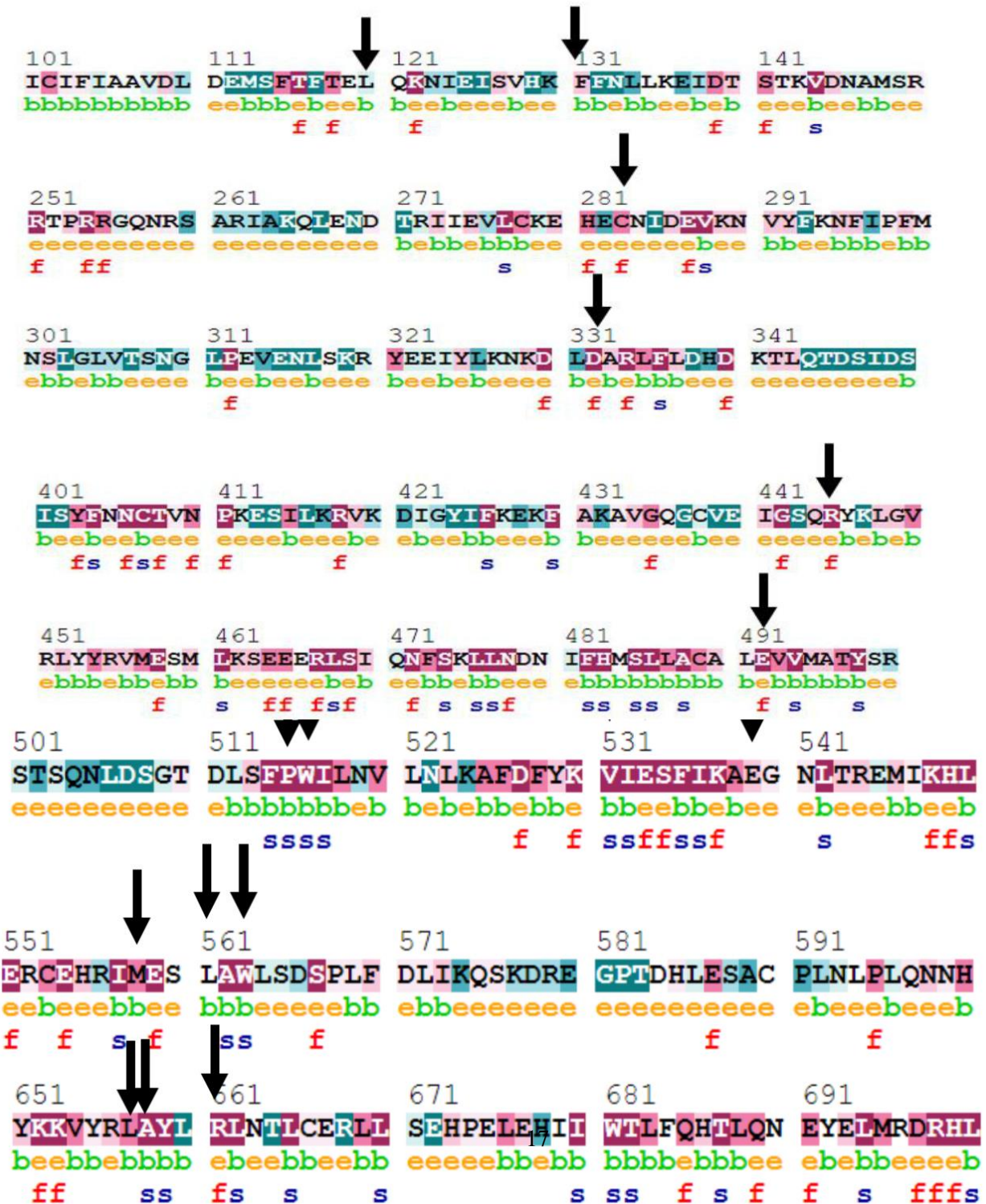

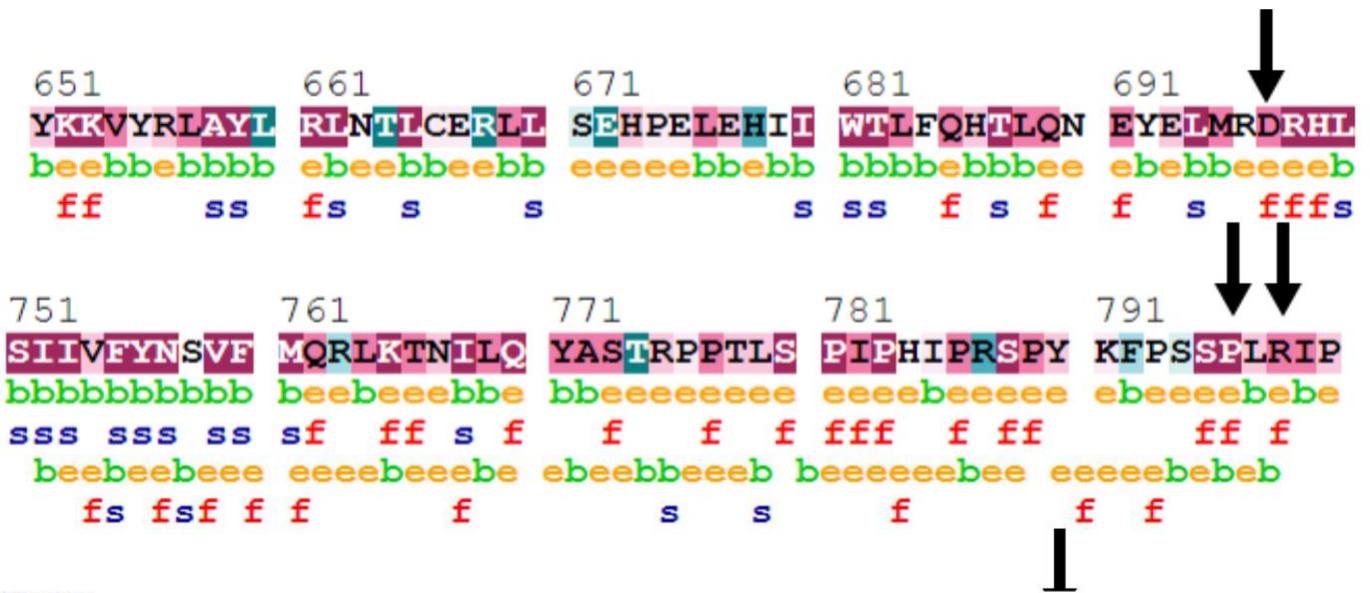

**Fig 2. The output of Consurf by the uniRef90 Protein database**

In consurf output, color shows the evolutionary conservation. Range from 1-4 shows variability

in Conservation is indicated by blue. Range from 5-6, shows average protection and

It is indicated by white. Range from 7-9 shows conservation and is characterized by purple. The residue (e)

Means exposed, (b) means buried,(f) means functional, i.e., highly conserved and exposed, (s) means

Structured, i.e., highly conserved and buried, and (x) means insufficient data. The black bold

Downward arrows represent

L120Q,F131L,C283Y,D332G,R445Q,E492V,P515T,W516G,V531G,E533K,E539K,M558RL561P

,W563G,L657Q,A658T,R661Q,D697H,D697E,P796L and R798W amino acid mutations of

SNPs i.e.

rs983885759,rs749495284,rs1273219762,rs763184576,rs747509282,rs771480219,rs866664638,rs138201027,rs143324585,rs1237070816,rs148379933,rs139494954,rs143400770,rs139500527,rs562956970,rs202119986,rs750578651,rs1358369644,rs3092903,rs1158706854 and rs187912365 respectively.)

From **Table 5**, **Residue (e) shows exposed, (b) shows buried,(f) shows functional i.e. highly conserved and exposed, (s) shows structured, i.e., highly conserved and buried**

Conservation scores from 1-4 means variable, 5-6 means intermediate, and 7-9 mean highly.

Conserved scores.

**Table 5:CONSERVATION PROFILE OF High Risk nsSNPs in Rb1 by UniRef-90**

| SR # | Rsid's | Residue & Position | Conservation Scores | Function |
|---|---|---|---|---|
| 1. | rs983885759 | L120Q | 4 | B |
| 2. | rs749495284 | F131L | 7 | B |
| 3. | rs1273219762 | C283Y | 8 | F |
| 4. | rs763184576 | D332G | 9 | F |
| 5. | rs747509282 | R445Q | 9 | F |
| 6. | rs771480219 | E492V | 9 | F |
| 7. | rs866664638 | P515T | 9 | S |
| 8. | rs138201027 | W516G | 9 | S |
| 9. | rs143324585 | V531G | 9 | S |
| 10. | rs1237070816 | E533K | 9 | F |
| 11. | rs148379933 | E539K | 7 | E |
| 12. | rs139494954 | M558R | 8 | B |
| 13. | rs143400770 | L561P | 5 | B |
| 14. | rs139500527 | W563G | 9 | S |
| 15. | rs562956970 | L657Q | 8 | B |
| 16. | rs202119986 | A658T | 9 | B |
| 17. | rs750578651 | R661Q | 9 | F |
| 18. | rs1358369644 | D697H | 8 | F |
| 19. | rs3092903 | D697E | 8 | F |
| 20. | rs1158706854 | P796L | 9 | F |
| 21. | rs187912365 | R798W | 8 | F |

From **Fig** and **"S3 Appendix"**, results can be concluded that out of **21** highly-risk diseased SNPs**, 19** are located on conserved regions(Score:7-9) while remaining two mutations i.e. L120Q, L561P Having scored 4 and 5, respectively, are located in non-conserved regions. Appendix S3, the The conservation score of conserved mutations is shown in **bold.** This means that these 19 Conserved mutations are much more damaging to the function and structure of the Rb1 protein.

### 3.10 Results of Project Hope:

The 19 conserved mutations declared by the consurf Server were being submitted into project hope in Order to get its 3D Protein structure and the physical properties of amino acids.

Moreover, Hope concluded that the size of mutant residue is less as compared to Wild-type residue. in many SNPs having rsid rs749495284 , rs763184576 , rs747509282 , rs771480219 , rs138201027 , rs143324585 , rs139500527 , rs750578651 at position F131L , D332G , R445Q , E492V , W516G , V531G , W563G and R661Q respectively. On the other hand, in many SNPs i.e. rs1273219762, rs1237070816, rs148379933, rs139494954, rs562956970 , rs202119986 , rs1358369644 , rs3092903 , rs1158706854 , rs187912365 at positions C283Y , E533K , E539K , M558R, L657Q , A658T , D697H , D697E , P796L, R798W respectively , the size of mutant The residue is greater than that of native residue.

As far as charges are concerned, the amount on wild-type was –I've whereas the control on mutant. The residue was neutral on many SNPs, i.e., rs763184576, rs771480219, rs1358369644 at positions.

D332G, E492V and D697H respectively. On the other hand, the charge on Wild-type residue was +ive whereas neutral on mutant residue on SNPs, i.e., rs750578651 and rs187912365 R661Q and R798W, respectively. The charge on wild-type residue was +ive, whereas –ive on.

Mutant residue on SNP, i.e., rs1237070816 at position E533K. The charge on wild-type residue was –ive while it is +ive on mutant residue on SNP, i.e. rs148379933 at position E539K. Wild-type was

Having neutral charge while it was +ive on mutant ones on SNP, i.e., rs139494954 at position M558R.

As hydrophobicity is concerned, the hydrophobicity of wild-type residue is more than mutant. residue in many SNPs i.e. rs866664638, rs138201027, rs143324585, rs139494954, rs139500527, rs562956970 and rs202119986 at positions P515T, W516G, V531G, M558R, W563G, L657Q and A658T respectively. The hydrophobicity of mutant residue is more significant than wild-type in many SNPs i.e. rs771480219 and rs187912365 at positions E492V and R798W respectively.

Protein structures of such mutations, as well as a change of amino acid regarding each nsSNP, are Shown in **the Table 6**

### **Table 6: Results of Project Hope**

| SNP ID | 3D Structure | Amino Acid Change |
|---|---|---|
| **rs749495284** | 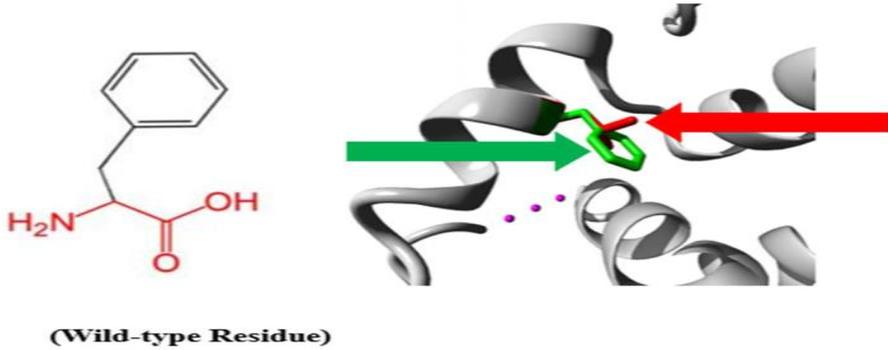 (Wild-type Residue) | **Phenylalanine Changed to Leucine at position 131.** |

| rs1273219762 | 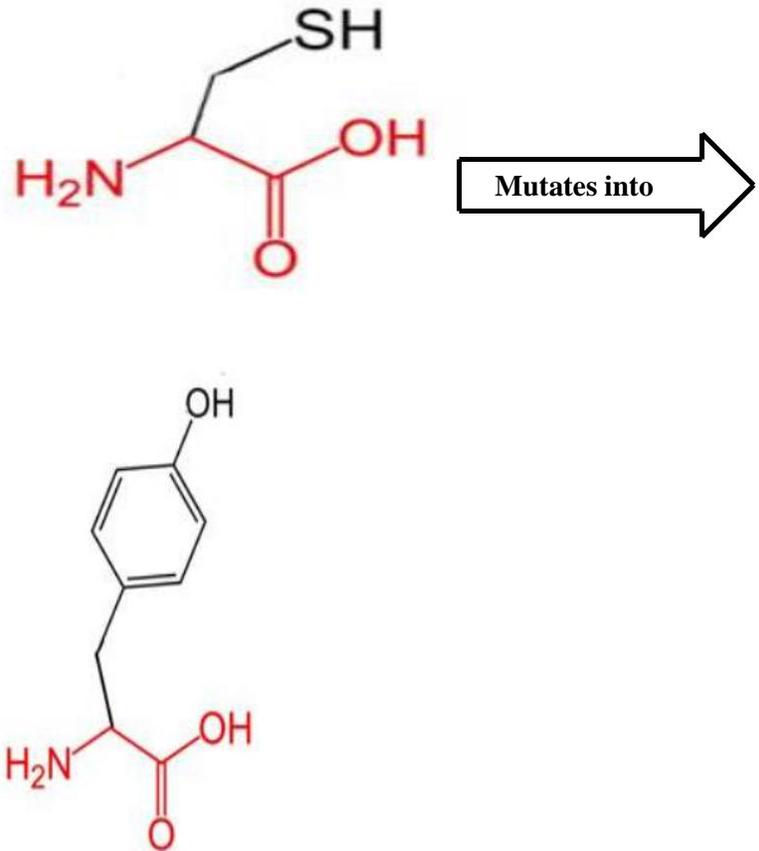 | Cysteine Changed to Tyrosine at position 283.<br><br>However the 3-D structure or modelling template is being missed on hope for unknown reason but administrator is being apprised in this regard. |
|---|---|---|
| rs763184576 | 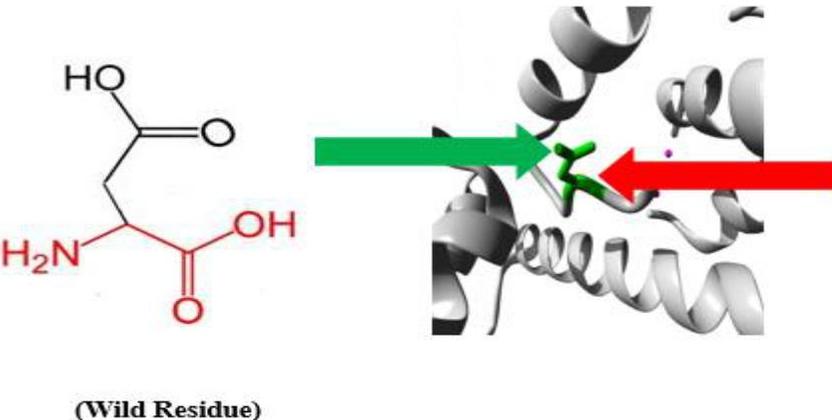<br>(Wild Residue) | Aspartic acid Changed to Glycine at position 332. |
| rs747509282 | 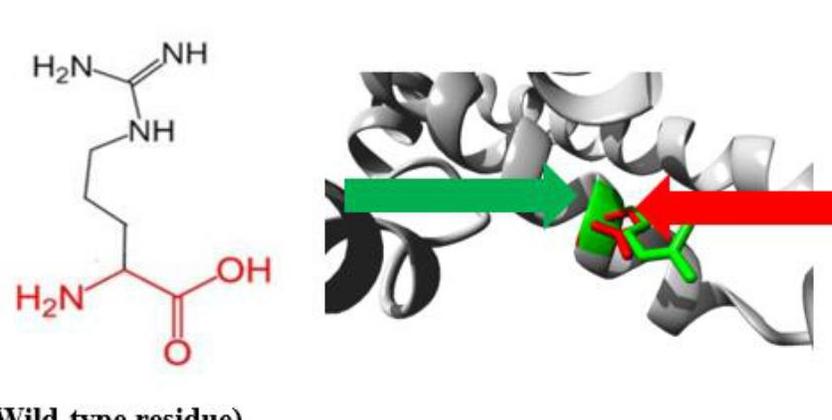<br>(Wild-type residue) | Arginine Changed to Glutamine at position 445. |

| rs771480219 | 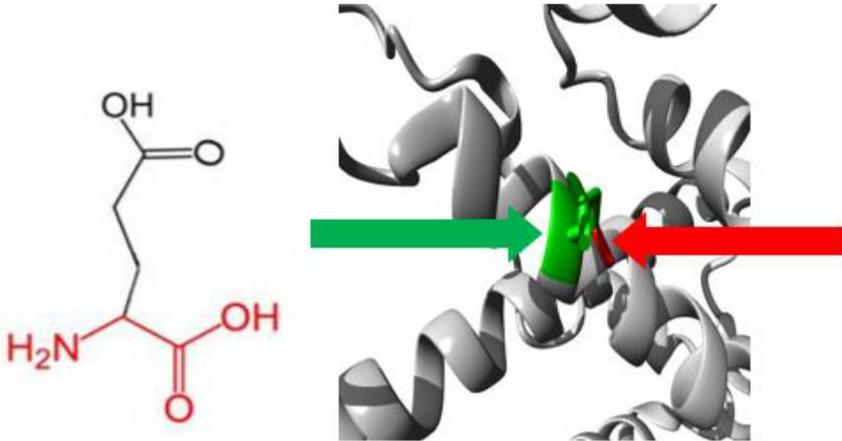 | **Glutamic Acid Changed to Valine at position 492.** |
| rs866664638 | 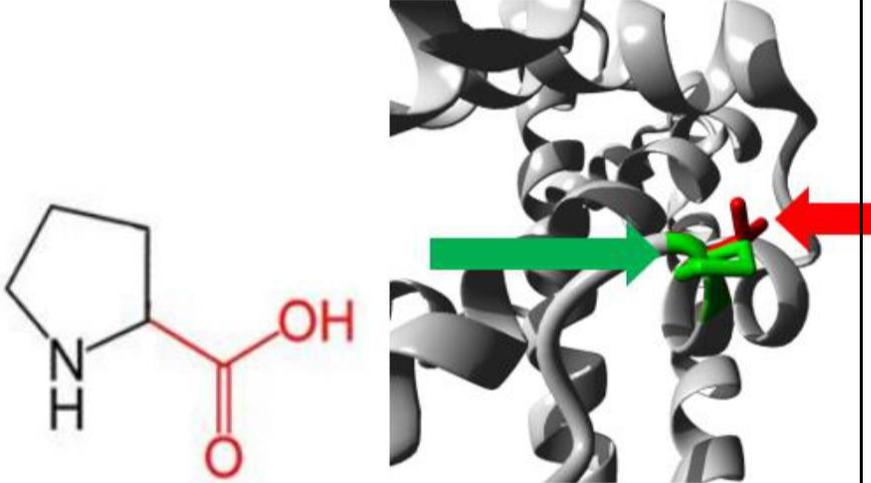 | **Proline Changed to Threonine at position 515.** |
| rs138201027 | 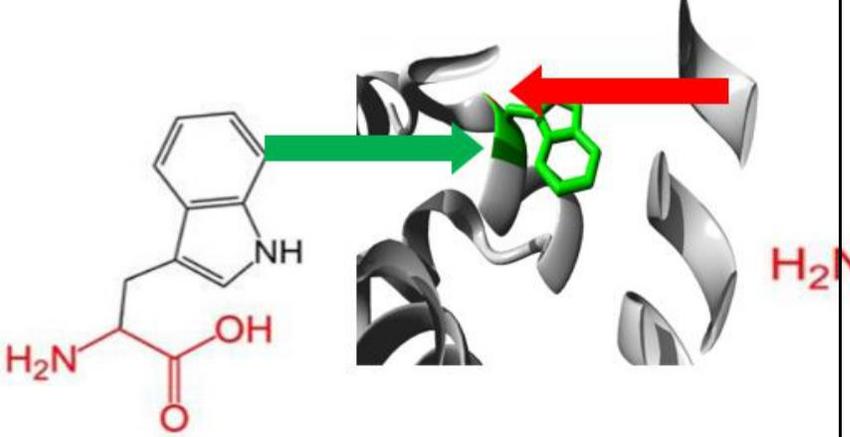 | **Tryptophan Changed to Glycin at position 516.** |

| rs143324585 | 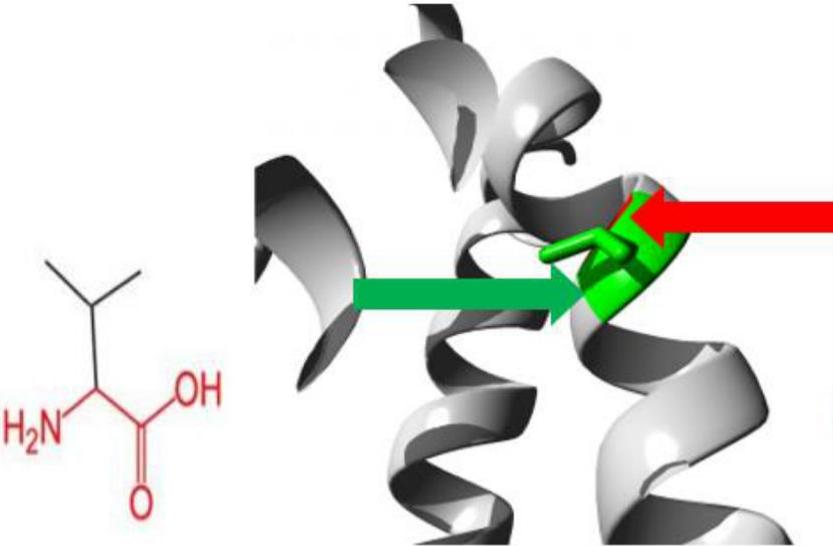 | Valine Changed to Glycine at position 531. |
| rs1237070816 | 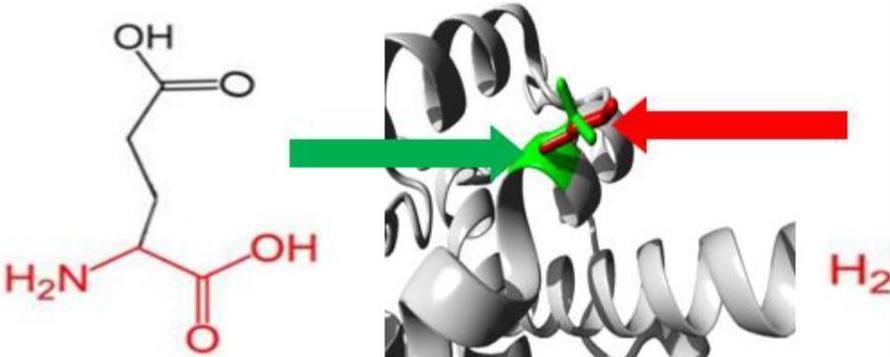 | Glutamic Acid changed to Lysine at position 533. |
| rs148379933 | 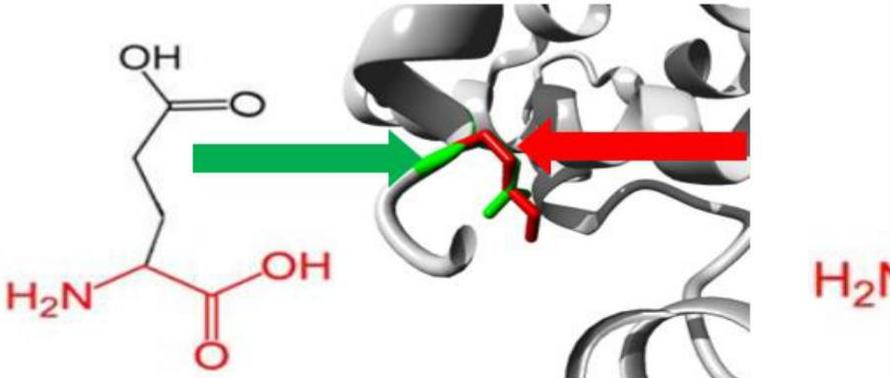 | Glutamic Acid changed to Lysine at position 539. |

| rs139494954 | 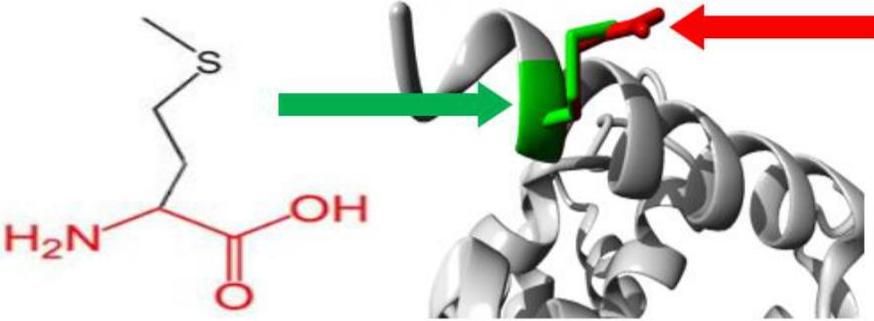 | Methionine Changed to Arginine at position 558. |
|---|---|---|
| rs139500527 | 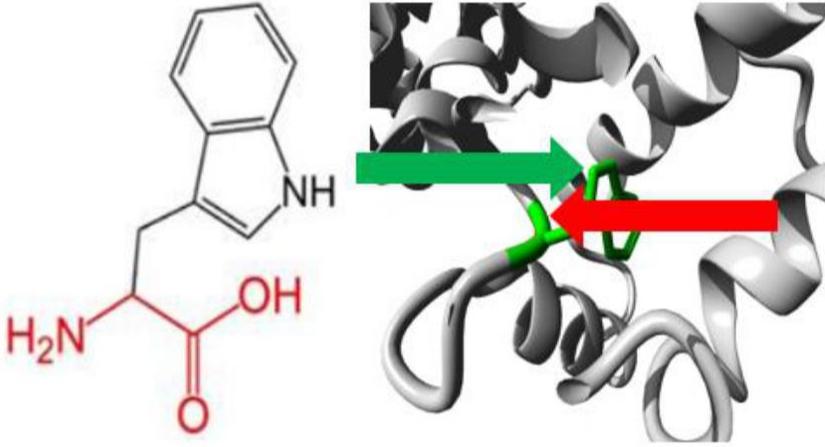 | Tryptophan Changed to Glycine at position 563. |
| rs562956970 | 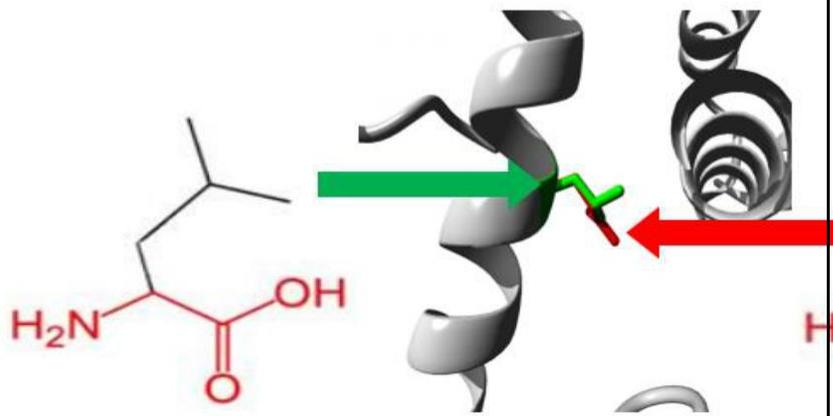 | Leucine Changed to Glutamine at position 657. |
| rs202119986 | 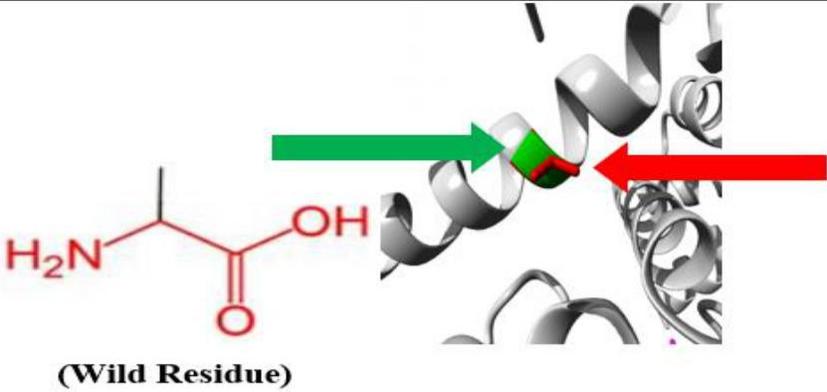 | Alanine mutated to Threonine at position 658. |
| rs750578651 | 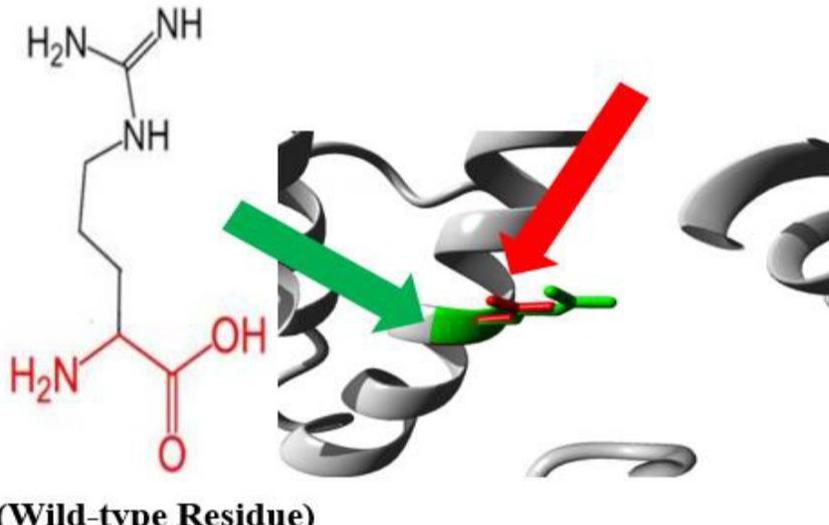 | Arginine Changed to Glutamine at position 661. |
| rs1358369644 | 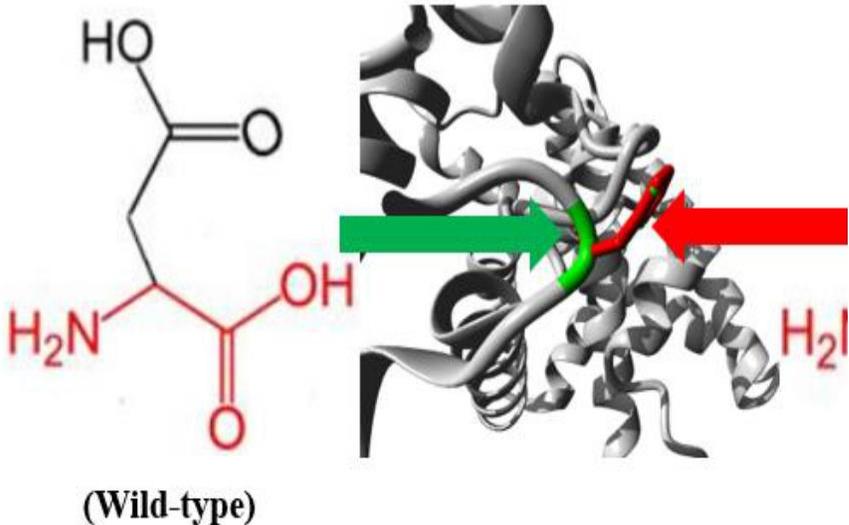 | Aspartic Acid Changed to Histidine at position 697. |

| rs3092903 | 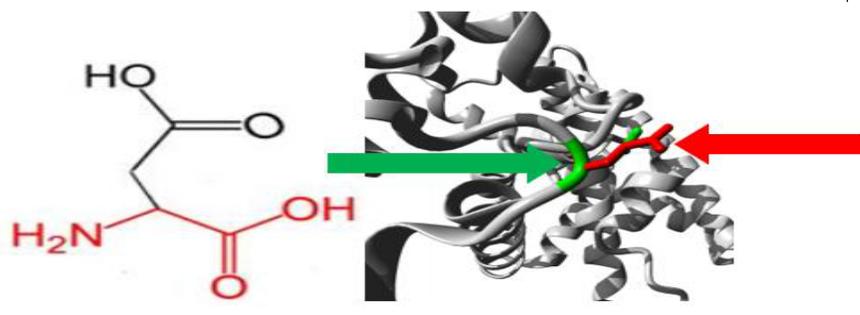 | Aspartic acid to Glutamic acid at position 697. |
|---|---|---|
| rs1158706854 | 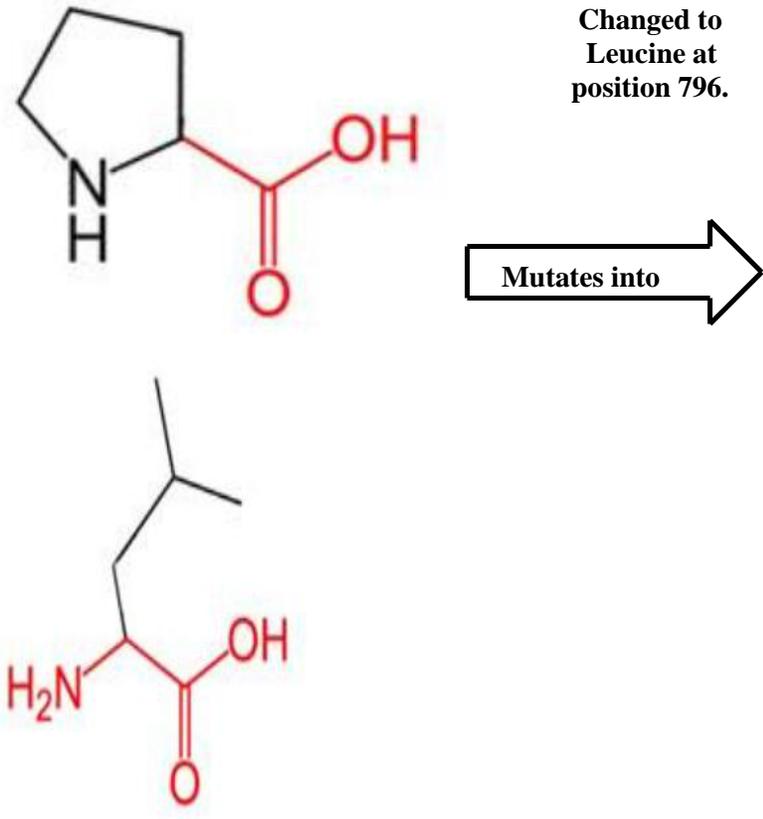  Proline Changed to Leucine at position 796.  Mutates into ⟹ | Protein structure can't be generated by hope because of lack of structural information. |

| rs187912365 | 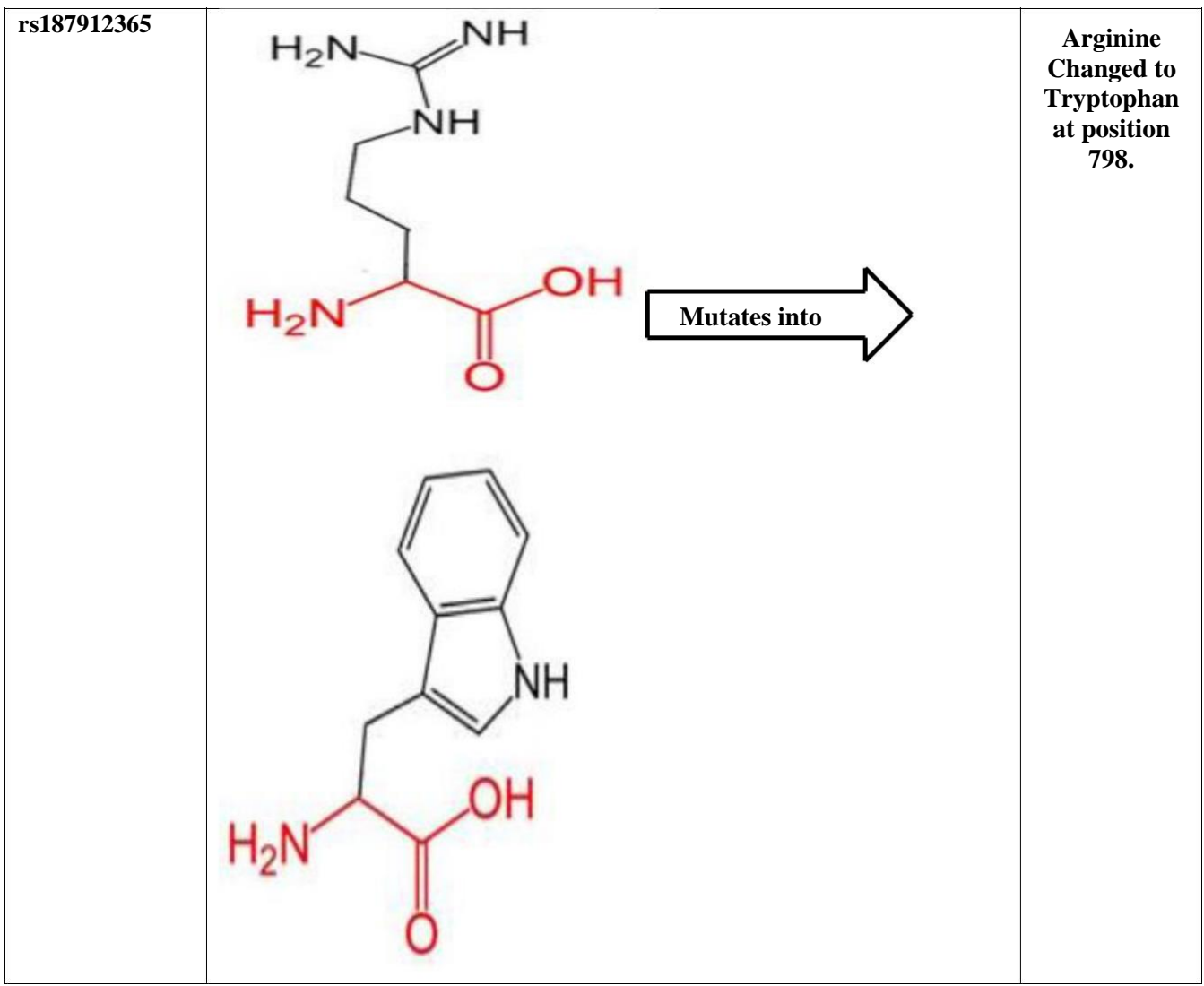 Mutates into | Arginine Changed to Tryptophan at position 798. |

Therefore, after giving protein structures of 19 mutations, Project Hope concluded that out of **19**, **17**

mutations i.e. D332G, R445Q, E492V, P515T, W516G, V531G, E533K, E539K, M558R, W563G, L657Q, A658T, R661Q, D697H, D697E, P796L and R798W are responsible for Rb1 protein

Damage. Just two mutations, i.e., F131L and C283Y, are not the sources of protein damage.

## 4 Discussion

The total number of recognized genomic variants, mainly SNPs, overgrowing in humans.Genome just because of highly advanced technologies. (22) The foremost goal of research in Population genetics and molecular biology are to point out deleterious SNPs from neutral ones. (23).

This study confirms the in-silico analysis of functional SNPs with the Rb1 gene.The novelty of this study is the isolation of highly deleterious SNPs from neutral ones with their protein Structures. Retinoblastoma (Rb1) is a tumor suppressor gene which means that its primary function is to control Cell growth from uneven growth prevents the cell from its rapid division. (24) However, the mutation in this gene causes Retinoblastoma, i.e., cancer of the retina in an eye or two, breast cancer,skin cancer, as well as lung cancer. Due to these severe disorders, it is mandatory to go to the root Cause of these mutations. (25)

This study undergoes a systematic approach to determine those functional SNPs in the human Rb1 gene and how that particular mutation is responsible for affecting the protein's structure and function and become the ultimate cause of many severe diseases. As we retrieved all data for the SNPs of the Rb1 Gene from dbSNP hosted by NCBA. There were a total of 36,358 SNPs. As the non-coding region is three hundred forty-five occurred in the 3'UTR region, 65 in 5'UTR, and 34,543 in the intron region. 844 SNPs were found in the coding region, out of which 199 were synonymous and 450 were non-synonymous which comprises 425 missense, five nonsense, and 20 frameshift mutations, and the remaining are other types of SNPs. We took only missense in his study as they directly affect the function and structure of Rb1 protein.

As we were supposed to compare the results of all the above-used tools and we found that out of 425, 40 were common deleterious SNPs declared deleterious by all the above tools i-e SIFT, Polyphen-2, I-Mutant 3.0, PROVEAN, and SNAP2. The next Step was to know whether these forty common deleterious SNPs were disease-causing polymorphisms or neutral ones. For this, we submitted our 40 common deleterious SNPs into "PHD-SNP, PMut, and SNPS&GO," and they declared that out of 40, 21 mutations were disease-causing. These 21 diseased mutations declared by PhD-SNP, PMut and SNPS&GO were submitted into the consurf Server to know which mutations are located in conserved regions as the mutations found in the conserved areas are much more deleterious than those SNPs located in non-conserved regions. So after submitting these 21 mutations into the consurf Server, it has been proved that out of 21, 19 diseased mutations were located in conserved regions. Then to get the protein structures of these 19 mutations as Well as to further verify whether these mutations are deleterious to the protein's function and structure or not, we submitted these 19 mutations to Project Hope and got their protein structure as well as related properties of amino acids.

After analyzing protein structures, Project Hope declared that out of 19, 17 mutations are the continuous source of protein damage. This 3D mutation analysis was done to find the location of nsSNPs and the variated properties of amino acids, i.e., hydrophobicity, charge, size, and flexibility.

Results of all the above-mentioned tools concluded that these 17 mutations are highly deleterious to the structure and function of the Rb1 gene, and these 17 mutations would help in the diagnosis and cure of all genetic diseases mentioned above, and just because of these flawless results, even the person suffering from Retinoblastoma, as well as other associated diseases, will take appropriate medication according to these mutations of his gene.

Shortly, any changes, whether minor or significant, in the Rb1 protein function it would affect the number of pathways involved in diseases.

# 5 Conclusion

Different bioinformatics tools have been used for the identification of harmful mutations which are the root cause of many severe diseases. All the data regarding SNPs, available on dbSNP hosted by NCBI has been utilized for analyzing deleterious SNPs as well as for evaluating their harmful effect on the structure and function of the Rb1 gene through various computational tools. Out of a total of 36,358 SNPs,345 were found in 3'UTR, 65 were in 5'UTR and 34,543 were in intronic region .844 were coding SNPs comprise 199 synonymous and 450 non-synonymous SNPs. Out of 450 non-Synonymous SNPs, 425 were missense, 05 were nonsense, and 20 were frameshift mutations and the remaining are other types of SNPs. For spotting the possible associations of the Rb1 gene with other genes, GeneMANIA, was being used. Although only 425 missenses were taken into account for further investigation. Ten different bioinformatics tools, i.e., Sift, Polyphen-2, I-Mutant 3.0, PROVEAN , SNAP2, PHD-SNP, PMut, SNPs&GO, Consurf and Project Hope have been used for flawless results as relying on only one tool, can't give accurate results. In the first step, 425 missense were passed from five tools i.e. Sift, Polyphen-2, I-Mutant 3.0, PROVEAN and SNAP2. In Second Step, only those SNPs declared damaging by all the above tools were passed from 4 other tools, i.e., Ph.D-SNP, PMut, SNP&GO, and SNP Analyzer to figure out whether these deleterious SNPs are disease-causing polymorphisms or neutral ones. As this is a matter of fact that those mutations found conserved regions tend to be more harmful than non-conserved regions. So It is much vital to find out those mutations found in conserved regions. So for this, in the 3[rd] Step, common disease-causing SNPs declared by the above tools were passed from Consurf Server to find out the evolutionarily conserved regions of SNPs. Then only those SNPs found in the conserved region were further passed through Project Hope to isolate those SNPs which are damaging to the structure and the Function of Rb1 protein and the protein structures of SNPs through Project Hope. Physical Properties of amino acids were also found.

Out of 425, after several iterative steps, only 17 SNPs were identified as damaging by all the above computational software.These are(F131L,C283Y,D332G,R445Q,E492V,P515T,W516G,V531G,E533K,E539K,M558R,W563G,L657Q,A658T,R661Q,D697H,D697E,P796L,R798W) mutations which directly29 affect the structure and function of the Rb1 gene become the ultimate cause of Retinoblastoma.These deleterious mutations mentioned above can be utilized for the diagnosis and cure of this worldwide and they are rapidly increasing grave issues, i.e., Retinoblastoma.

6. **References**

[1] T. W. Corson and B. L. Gallie, "One hit, two hits, three hits, more? Genomic changes in the development of retinoblastoma," *Genes Chromosomes and Cancer*, vol. 46, no. 7. Genes Chromosomes Cancer, pp. 617–634, Jul. 2007, doi: 10.1002/gcc.20457.

[2] J. Zhang *et al.*, "A novel retinoblastoma therapy from genomic and epigenetic analyses," *Nature*, vol. 481, no. 7381, pp. 329–334, Jan. 2012, doi: 10.1038/nature10733.

[3] A. G. Knudson, "Mutation and cancer: statistical study of retinoblastoma.," *Proc. Natl. Acad. Sci. U. S. A.*, vol. 68, no. 4, pp. 820–823, 1971, doi: 10.1073/pnas.68.4.820.

[4] K. G. Ewens *et al.*, "Phosphorylation of pRb: mechanism for RB pathway inactivation in MYCN-amplified retinoblastoma," *Cancer Med.*, vol. 6, no. 3, pp. 619–630, Mar. 2017, doi: 10.1002/cam4.1010.

[5] K. C. Sippel *et al.*, "Frequency of somatic and germ-line mosaicism in retinoblastoma: Implications for genetic counseling," *Am. J. Hum. Genet.*, vol. 62, no. 3, pp. 610–619, Mar. 1998, doi: 10.1086/301766.

[6] L. Su *et al.*, "Research on Single Nucleotide Polymorphisms Interaction Detection from Network Perspective," *PLoS One*, vol. 10, no. 3, p. e0119146, Mar. 2015, doi: 10.1371/journal.pone.0119146.

[7] S. W. Doniger *et al.*, "A Catalog of Neutral and Deleterious Polymorphism in Yeast," *PLoS Genet.*, vol. 4, no. 8, p. e1000183, Aug. 2008, doi: 10.1371/journal.pgen.1000183.

[8] J. Havlicek *et al.*, "An application of competitive reporter monitored amplification (CMA) for rapid detection of single nucleotide polymorphisms (SNPs)," *PLoS One*, vol. 12, no. 8, p. e0183561, Aug. 2017, doi: 10.1371/journal.pone.0183561.

[9] Z. Li *et al.*, "Prediction and Analysis of Retinoblastoma Related Genes through Gene Ontology and KEGG," *Biomed Res. Int.*, vol. 2013, 2013, doi: 10.1155/2013/304029.

[10] L. S. Mian, A. Aqil, K. Abid, and M. Moin, "Outcome of interventional treatment modalities for retinoblastoma: Experience at a tertiary care centre in Pakistan," *J. Pak. Med. Assoc.*, vol. 69, no. 7, pp. 1039–1043, 2019.

[11] J. Madhavan, A. Ganesh, and G. Kumaramanickavel, "Retinoblastoma: From disease to discovery," *Ophthalmic Res.*, vol. 40, no. 5, pp. 221–226, 2008, doi: 10.1159/000128578.

[12] A. MacCarthy *et al.*, "Retinoblastoma in Great Britain 1963-2002," *Br. J. Ophthalmol.*, vol. 93, no. 1, pp. 33–37, 2009, doi: 10.1136/bjo.2008.139618.

[13] A. Melamud, R. Palakar, and A. Singh, "Retinoblastoma," *Am. Fam. Physician*, vol. 73, no. 6, pp. 1039–1044, Mar. 2006.

[14] S. E. Soliman, H. Racher, C. Zhang, H. MacDonald, and B. L. Gallie, "Genetics and molecular diagnostics in retinoblastoma - An Update," *Asia-Pacific Journal of Ophthalmology*, vol. 6, no. 2. Asia-Pacific Academy of Ophthalmology, pp. 197–207, 2017, doi: 10.22608/APO.201711.

[15] D. R. Lohmann and B. L. Gallie, "Retinoblastoma: Revisiting the model prototype of inherited cancer," *American Journal of Medical Genetics - Seminars in Medical Genetics*, vol. 129 C, no. 1. Wiley-Liss Inc., pp. 23–28, Aug. 15, 2004, doi: 10.1002/ajmg.c.30024.

[16] D. Parma, M. Ferrer, L. Luce, F. Giliberto, and I. Szijan, "RB1 gene mutations in Argentine retinoblastoma patients. Implications for genetic counseling," *PLoS One*, vol. 12, no. 12, p. e0189736, Dec. 2017, doi: 10.1371/journal.pone.0189736.

[17] M. A. Gener *et al.*, "Clinical, Pathological, and Surgical Outcomes for Adult Pineoblastomas," *World Neurosurg.*, vol. 84, no. 6, pp. 1816–1824, 2015, doi: 10.1016/j.wneu.2015.08.005.

[18] R. A. Kleinerman, S. J. Schonfeld, and M. A. Tucker, "Sarcomas in hereditary retinoblastoma," *Clin. Sarcoma Res.*, vol. 2, no. 1, 2012, doi: 10.1186/2045-3329-2-15.

[19] S. R. Freeman *et al.*, "Statins, fibrates, and melanoma risk: A systematic review and meta-analysis," *J. Natl. Cancer Inst.*, vol. 98, no. 21, pp. 1538–1546, 2006, doi: 10.1093/jnci/djj412.

[20] M. Sanlorenzo *et al.*, "The risk of melanoma in airline pilots and cabin crew: A meta-analysis," *JAMA Dermatology*, vol. 151, no. 1, pp. 51–58, Jan. 2015, doi: 10.1001/jamadermatol.2014.1077.

[21] A. C. Schefler and D. H. Abramson, "Retinoblastoma: What is new in 2007-2008," *Curr. Opin. Ophthalmol.*, vol. 19, no. 6, pp. 526–534, 2008, doi: 10.1097/ICU.0b013e328312975b.

[22] R. Rajasekaran, C. Sudandiradoss, C. George, P. Doss, and R. Sethumadhavan, "Identification and in silico analysis of functional SNPs of the BRCA1 gene," *Genomics*, vol. 90, pp. 447–452, 2007, doi: 10.1016/j.ygeno.2007.07.004.

[23] C. George Priya Doss and S. Rao, "Impact of single nucleotide polymorphisms in HBB gene causing haemoglobinopathies: in silico analysis," *N. Biotechnol.*, vol. 25, no. 4, pp. 214–219, 2009, doi: 10.1016/j.nbt.2009.01.004.

[24] R. Rajasekaran and R. Sethumadhavan, "Exploring the structural and functional effect of pRB by significant nsSNP in the coding region of RB1 gene causing retinoblastoma," *Sci. China Life Sci.*, vol. 53, no. 2, pp. 234–240, 2010, doi: 10.1007/s11427-010-0039-y.

[25] M. Alanazi *et al.*, "In silico analysis of single nucleotide polymorphism (snps) in human β-globin gene," *PLoS One*, vol. 6, no. 10, 2011, doi: 10.1371/journal.pone.0025876.

[26] M. R. M. Hussain *et al.*, "In silico analysis of Single Nucleotide Polymorphisms (SNPs) in human BRAF gene," *Gene*, vol. 508, no. 2, pp. 188–196, 2012, doi: 10.1016/j.gene.2012.07.014.

[27]  S. Rajamanikandan, R. Vanajothi, A. Sudha, P. Rameshthangam, and P. Srinivasan, "In silico analysis of deleterious SNPs of the FGFR2 gene," *J. Biol. Sci.*, vol. 12, no. 2, pp. 83–90, 2012, doi: 10.3923/jbs.2012.83.90.

[28]  B. Dabhi and K. N. Mistry, "In silico analysis of single nucleotide polymorphism (SNP) in human TNF-α gene," *Meta Gene*, vol. 2, pp. 586–595, Dec. 2014, doi: 10.1016/j.mgene.2014.07.005.

[29]  M. Jia, B. Yang, Z. Li, H. Shen, X. Song, and W. Gu, "Computational analysis of functional single nucleotide polymorphisms associated with the CYP11B2 gene," *PLoS One*, vol. 9, no. 8, Aug. 2014, doi: 10.1371/journal.pone.0104311.

[30]  H. Sheikh *et al.*, "Research Article First Comprehensive In Silico analysis of the functional and structural consequences of SNPs in human GalNAc-T1 gene.," *Comput. Math. Methods Med.*, 2014, doi: 10.1155/2014/904052.

[31]  V. Shanthi, R. Rajasekaran, and K. Ramanathan, "Computational identification of significant missense mutations in AKT1 gene," *Cell Biochem. Biophys.*, vol. 70, no. 2, pp. 957–965, Nov. 2014, doi: 10.1007/s12013-014-0003-8.

[32]  F. R. Marín-Martín, C. Soler-Rivas, R. Martín-Hernández, and A. Rodriguez-Casado, "A Comprehensive In Silico Analysis of the Functional and Structural Impact of Nonsynonymous SNPs in the ABCA1 Transporter Gene," *Hindawi Publ. Corp. Cholest.*, p. 20, 2014, doi: 10.1155/2014/639751.

[33]  M. Noreen, S. Murad, and H. Khan, "In silico analysis of SNPs in coding region of human c-Myc gene," *Pak. J. Zool.*, vol. 47, no. 5, pp. 1305–1318, 2015.

[34]  Z. Mahmud, S. U. F. Malik, J. Ahmed, and A. K. Azad, "Computational analysis of damaging single-nucleotide polymorphisms and their structural and functional impact on the insulin receptor," *Biomed Res. Int.*, vol. 2016, 2016, doi: 10.1155/2016/2023803.

[35]  M. Mohamed Osman Saudi German Hospital, S. Arabia, and H. N. Altayb, "In silico Analysis of Single Nucleotide Polymorphisms (Snps) in Human FTO Gene," 2016. Accessed: Jul. 01, 2020. [Online]. Available: https://www.researchgate.net/publication/301801114.

[36]  N. E. Abdelraheem, G. M. El-tayeb, L. O. Osman, and S. A. Abedlrhman, "A comprehensive in silico analysis of the functional and structural impact of non-synonymous single nucleotide polymorphisms in the human KRAS gene," *Am. J. Bioinforma. Res.*, vol. 6, no. 2, pp. 32–55, 2016, doi: 10.5923/j.bioinformatics.20160602.02.

[37]   A. Abd Elhamid Fadlalla Elshaikh *et al.*, "Computational Analysis of Single Nucleotide Polymorphism (SNPs) in Human GRM4 Gene," *Am. J. Biomed. Res.*, vol. 4, no. 3, pp. 61–73, 2016, doi: 10.12691/ajbr-4-3-2.

[38]   Z. Kosaloglu *et al.*, "In silico SNP analysis of the breast cancer antigen NY-BR-1," *BMC Cancer*, vol. 16, no. 1, p. 901, Nov. 2016, doi: 10.1186/s12885-016-2924-7.

[39]   F. Abdul Samad, B. A. Suliman, S. H. Basha, T. Manivasagam, and M. M. Essa, "A Comprehensive In Silico Analysis on the Structural and Functional Impact of SNPs in the Congenital Heart Defects Associated with NKX2-5 Gene—A Molecular Dynamic Simulation Approach," *PLoS One*, vol. 11, no. 5, p. e0153999, May 2016, doi: 10.1371/journal.pone.0153999.

[40]   M. Desai and J. Chauhan, "In silico analysis of nsSNPs in human methyl CpG binding protein 2," *Meta Gene*, vol. 10, pp. 1–7, Dec. 2016, doi: 10.1016/j.mgene.2016.09.004.

[41]   M. Nimir *et al.*, "In silico analysis of single nucleotide polymorphisms (SNPs) in human FOXC2 gene," *F1000Research*, vol. 6, no. January, 2017, doi: 10.12688/f1000research.10937.2.

[42]   K. Renu *et al.*, "In silico Analysis of Functional Single Nucleotide Polymorphisms in Genes Related to Adipose Tissue Impairment," 2017. Accessed: Jun. 29, 2020. [Online]. Available: http://genetics.

[43]   M. Solayman, M. A. Saleh, S. Paul, M. I. Khalil, and S. H. Gan, "In silico analysis of nonsynonymous single nucleotide polymorphisms of the human adiponectin receptor 2 (ADIPOR2) gene," *Comput. Biol. Chem.*, vol. 68, pp. 175–185, 2017, doi: 10.1016/j.compbiolchem.2017.03.005.

[44]   A. Elnasri, A. M. Al Bkrye, and M. A. M. Khaier, "In Silico Analysis of Non Synonymous SNPs in DHCR7 Gene," *Am. J. Bioinforma. Res.*, vol. 8, no. 1, pp. 12–18, 2018, doi: 10.5923/j.bioinformatics.20180801.02.

[45]   A. Arifuzzaman *et al.*, "In Silico Analysis of Non Synonymous Single Nucleotide Polymorphisms (nsSNPs) of SMPX Gene in Hearing Impairment," *bioRxiv Bioinforma.*, p. 461764, Nov. 2018, doi: 10.1101/461764.

[46]   S. Elbager *et al.*, "Computational Analysis of Deleterious Single Nucleotide Polymorphisms (SNPs) in Human CALR Gene Diagnostic Parasitology View project Myeloproliferative Neoplasms in Sudan View project Computational Analysis of Deleterious Single Nucleotide Polymorphisms (," *Am. J. Bioinforma. Res.*, vol. 8, no. 1, pp. 1–11, 2018, doi: 10.5923/j.bioinformatics.20180801.01.

[47]   M. Arshad, A. Bhatti, and P. John, "Identification and in silico analysis of functional SNPs of human TAGAP protein: A comprehensive study," *PLoS One*, vol. 13, no. 1, p. e0188143, Jan. 2018, doi: 10.1371/journal.pone.0188143.

[48]    L. Elkhattabi *et al.*, "In silico analysis of coding/noncoding SNPs of human RETN gene and characterization of their impact on resistin stability and structure," *J. Diabetes Res.*, vol. 2019, 2019, doi: 10.1155/2019/4951627.

[49]    S. Mostafavi, D. Ray, D. Warde-Farley, C. Grouios, and Q. Morris, "GeneMANIA: A real-time multiple association network integration algorithm for predicting gene function," *Genome Biol.*, vol. 9, no. SUPPL. 1, p. S4, Jun. 2008, doi: 10.1186/gb-2008-9-s1-s4.

[50]    S. A. De Alencar and J. C. D. Lopes, "A Comprehensive In Silico Analysis of the Functional and Structural Impact of SNPs in the IGF1R Gene," *J. Biomed. Biotechnol.*, vol. 2010, 2010, doi: 10.1155/2010/715139.

[51]    S. M. O. Sarour *et al.*, "New mutation found within OTOR gene involved in deafness in two Sudanese families from Al-Jazirah state- Sudan : using Next Generation Sequencing ( NGS )," *Bio-Genetics J.*, vol. 2, no. 6, pp. 46–50, 2014.

[52]    I. Adzhubei, D. M. Jordan, and S. R. Sunyaev, "Predicting functional effect of human missense mutations using PolyPhen-2," *Curr. Protoc. Hum. Genet.*, vol. 0 7, no. SUPPL.76, p. Unit7.20, 2013, doi: 10.1002/0471142905.hg0720s76.

[53]    A. Datta, M. Habibul Hasan Mazumder, A. Sultana Chowdhury, and M. Anayet Hasan, "Functional and Structural Consequences of Damaging Single Nucleotide Polymorphisms in Human Prostate Cancer Predisposition Gene RNASEL," *Biomed Res. Int.*, 2015, doi: 10.1155/2015/271458.

[54]    E. Capriotti, P. Fariselli, and R. Casadio, "I-Mutant2.0: predicting stability changes upon mutation from the protein sequence or structure," *Nucleic Acids Res.*, 2005, doi: 10.1093/nar/gki375.

[55]    T. G. Kucukkal, Y. Yang, S. C. Chapman, W. Cao, and E. Alexov, "Computational and experimental approaches to reveal the effects of single nucleotide polymorphisms with respect to disease diagnostics," *International Journal of Molecular Sciences*, vol. 15, no. 6. MDPI AG, pp. 9670–9717, May 30, 2014, doi: 10.3390/ijms15069670.

[56]    M. Hecht, Y. Bromberg, and B. Rost, "Better prediction of functional effects for sequence variants," *BMC Genomics*, vol. 16, no. 8, p. S1, Jun. 2015, doi: 10.1186/1471-2164-16-S8-S1.

[57]    E. Capriotti and P. Fariselli, "PhD-SNP g : a webserver and lightweight tool for scoring single nucleotide variants," *Nucleic Acids Res.*, vol. 45, pp. 247–252, 2017, doi: 10.1093/nar/gkx369.

[58]    M. Seifi and M. A. Walter, "Accurate prediction of functional, structural, and stability changes in PITX2 mutations using in silico bioinformatics algorithms," *PLoS One*, vol. 13, no. 4, p. e0195971, Apr. 2018, doi: 10.1371/journal.pone.0195971.

[59] M. M. Merghani *et al.*, "In silico analysis of Single Nucleotide Polymorphisms (SNPs) in Human vWF Gene," 2017. Accessed: Jun. 28, 2020. [Online]. Available: http://snps-and-go.biocomp.unibo.it/snps-and-go/.

[60] F. Glaser *et al.*, "ConSurf: Identification of Functional Regions in Proteins by Surface-Mapping of Phylogenetic Information," Valdar and Thornton, 2003. Accessed: Jun. 28, 2020. [Online]. Available: http://consurf.tau.ac.il.

[61] I. Mayrose, D. Graur, N. Ben-Tal, and T. Pupko, "Comparison of site-specific rate-inference methods for protein sequences: Empirical Bayesian methods are superior," *Mol. Biol. Evol.*, vol. 21, no. 9, pp. 1781–1791, 2004, doi: 10.1093/molbev/msh194.

[62] H. Venselaar, T. A. H. te Beek, R. K. P. Kuipers, M. L. Hekkelman, and G. Vriend, "Protein structure analysis of mutations causing inheritable diseases. An e-Science approach with life scientist friendly interfaces," *BMC Bioinformatics*, vol. 11, Nov. 2010, doi: 10.1186/1471-2105-11-548.

[63] S. T. Sherry *et al.*, "dbSNP: the NCBI database of genetic variation," 2001. Accessed: Jun. 29, 2020. [Online]. Available: http://www.ncbi.nlm.nih.gov/SNP.

[64] M. Zhu and S. Zhao, "Candidate gene identification approach: Progress and challenges," *International Journal of Biological Sciences*, vol. 3, no. 7. Ivyspring International Publisher, pp. 420–427, Oct. 25, 2007, doi: 10.7150/ijbs.3.420.

[65] E. A. Price *et al.*, "Spectrum of RB1 mutations identified in 403 retinoblastoma patients," *J. Med. Genet.*, vol. 51, no. 3, pp. 208–214, 2014, doi: 10.1136/jmedgenet-2013-101821.

[66] G. De Falco and A. Giordano, "pRb2/p130: A new candidate for retinoblastoma tumor formation," *Oncogene*, vol. 25, no. 38. Oncogene, pp. 5333–5340, Aug. 28, 2006, doi: 10.1038/sj.onc.1209614.